\documentclass{ieeetj}
\usepackage{cite}
\usepackage{amsmath,amssymb,amsfonts}
\usepackage{graphicx,color}
\usepackage{textcomp}
\usepackage{xcolor}
\usepackage{hyperref}
\hypersetup{hidelinks=true}
\usepackage{algorithm,algorithmic}
\def\BibTeX{{\rm B\kern-.05em{\sc i\kern-.025em b}\kern-.08em
    T\kern-.1667em\lower.7ex\hbox{E}\kern-.125emX}}
\AtBeginDocument{\definecolor{tmlcncolor}{cmyk}{0.93,0.59,0.15,0.02}\definecolor{NavyBlue}{RGB}{0,86,125}}

\def\OJlogo{\vspace{-4pt}$<$Society logo(s) and publication title will appear here.$>$}
\def\seclogo{\vspace{10pt}$<$Society logo(s) and publication title will appear here.$>$}

\def\authorrefmark#1{\ensuremath{^{\textbf{#1}}}}

\begin{document}

\newtheorem{Thm}{Theorem}[section]
\newtheorem{Cor}[Thm]{Corollary}
\newtheorem{Lem}[Thm]{Lemma}
\newtheorem{Prop}[Thm]{Proposition}
\newtheorem{Def}[Thm]{Definition}
\newtheorem{Assump}[Thm]{Assumption}
\newtheorem{Rem}[Thm]{Remark}
\newtheorem{Exam}[Thm]{Example}

\receiveddate{XX Month, XXXX}
\reviseddate{XX Month, XXXX}
\accepteddate{XX Month, XXXX}
\publisheddate{XX Month, XXXX}
\currentdate{XX Month, XXXX}
\doiinfo{XXXX.2022.1234567}

\markboth{Dynamic Orthogonal Matching Pursuit for Sparse Data Reconstruction}{ZHAO and LUO}

\title{Dynamic Orthogonal Matching Pursuit for Sparse Data Reconstruction}

\author{Yun-Bin Zhao\authorrefmark{1}, Member, IEEE, and  Zhi-Quan Luo\authorrefmark{1}, Fellow, IEEE}
\affil{Shenzhen Research Institute of Big Data, Chinese University of Hong Kong, Shenzhen, Guangdong Province, China}
\corresp{Corresponding author: Yun-Bin Zhao (email: yunbinzhao@cuhk.edu.cn).}
\authornote{This work was supported in part by the National Natural Science Foundation of China (NSFC)  under grants 12071307, 61571384 and 61731018.}

\begin{abstract}
The orthogonal matching pursuit (OMP) is one of the mainstream algorithms for sparse data reconstruction or approximation.  It acts as a driving force for the development of several other greedy methods for sparse data reconstruction, and it also plays a vital role in the development of compressed sensing theory for sparse signal and image reconstruction. In this paper, we propose the so-called  dynamic orthogonal matching pursuit (DOMP) and enhanced dynamic orthogonal matching pursuit (EDOMP) algorithms which are more efficient than OMP for sparse data reconstruction from a numerical point of view.  We carry out a rigorous analysis to establish the reconstruction error bound for  DOMP under the restricted isometry property of the measurement matrix. The main result claims that the reconstruction error via   DOMP  can be controlled  and  measured in terms of the number of iterations,   sparsity level of data, and the noise level of measurements. Moveover, the finite convergence of DOMP for a class of large-scale compressed sensing problems is also shown.
\end{abstract}


\maketitle

\section{Introduction} \label{section1}
 \IEEEPARstart{D}{ata} reconstruction/recovery is a common request in such areas as signal denoising, imaging reconstruction, statistical model selection, pattern recognition, streaming data tracking and wireless  channel estimation (see, e.g., \cite{E10,EK12,GDHIWH18,BCKV19,LCG19}). For instance, recovering a sparse signal from limited measurements (observations) is a central task in compressed-sensing-based signal processing \cite{E10,EK12,CRT06,D06,FR13}.  Thus developing fast and robust algorithms for data reconstruction is significantly important in these  scenarios. Without loss of generality, let $x\in \mathbb{R}^n$ denote the data (e.g., signal or image) which admits some sparsity structure in the sense that $x$ is either $k$-sparse (i.e., $x$ has at most $k\ll n $ nonzero components) or $k$-compressible (i.e., $ x$ can be approximated by a $k$-sparse vector). In this paper, the unknown data $ x\in \mathbb{R}^n$  to reconstruct is called the target data. To reconstruct the data $x\in \mathbb{R}^n, $   one may first collect a few measurements  \begin{equation} \label{measurement} y_i: = (a^i)^T x +\nu_i , ~   i=1, \cdots, m, \end{equation}  where $ m \ll n, $  $a^i$'s are given measurement vectors, and $ \nu_i$'s denote the measurement errors. Let $A$ be the $ m\times n$ matrix whose row vectors consist of $(a^i)^T,   i=1, \dots, m,$ and let $  y = (y_1, \dots, y_m)^T $ and $ \nu= (\nu_1, \dots, \nu_m)^T $ be the vectors of   measurements and   errors, respectively. Then the system (\ref{measurement}) can be written as $  y: = A x + \nu . $
Under the assumption that the target data $x$ is $k$-sparse or $k$-compressible, the reconstruction of $x\in \mathbb{R}^n $ from the measurements $y$ can be formulated as the sparse optimization problem
\begin{equation} \label{L0} \min_{z\in \mathbb{R}^n} \{\|y- Az\|^2_2: ~ \|z\|_0 \leq k\}, \end{equation}  where $ \|z\|_0 $ denotes the number of nonzero entries of $z.$
Essentially, the model (\ref{L0}) is a combinatorial optimization problem. The vectors satisfying the constraint of (\ref{L0}) are $k$-sparse. In this paper, we assume that $ k\ll n$ which is a typical assumption in the scenario of compressed-sensing-based signal and image reconstruction \cite{E10,EK12,FR13,MEH19}.  The algorithms for (\ref{L0}) and similar problems have experienced a significant development over the past decades. The most widely used  algorithms are based on convex/nonconvex optimization (e.g., \cite{CDS98,CT05,CWB08,BT09,BE13b,Z18}), hard thresholding and  greedy methods (e.g., \cite{E10,EK12,FR13,KK18}), and the recent technique based on the concept of optimal $k$-thresholding \cite{Z20, ZL21, MZ20}.  The orthogonal matching pursuit (OMP) is one of the popular greedy methods for the problem (\ref{L0}).
Before recalling such a method, let us first introduce some notations used in the paper.

\subsection{Notation}
Throughout the paper, all vectors are column vectors unless otherwise specified.
The transpose of the vector $z$ and matrix $A$ are denoted by $ z^T$ and $A^T,$ respectively. The support of the vector $z$ is denoted by $ \textrm{supp} ( z) = \{i: z_i \not=0\}. $  We use $ {\cal H}_k(\cdot)$ to denote the hard thresholding operator which retains the $k$ largest entries in magnitude and zeros out the other entries of a vector, and we use $L_q(\cdot) $ to denote the index set of the $q$  largest entries in  magnitude of a vector.  When two entries of $z$ are equal in absolute value, ${\cal H}_k(z)$ and $L_q (z)$ might not be uniquely determined, in which case we select the entry with the smallest index.  The complement set of  $ S \subseteq \{1, 2, \dots, n\} $ with respect to $\{1, \dots, n\}$ is denoted by $ \overline{S}=  \{1, 2, \dots, n\} \backslash S , $ and the cardinality of $ S $  is denoted by $|S|  .$    Given $ x\in \mathbb{R}^n$ and $ S\subseteq \{1, 2, \dots, n\},$ the vector $ x_S \in \mathbb{R}^n$ denotes the vector obtained from $ x$ by retaining the entries of $ x$ supported on $S$ and setting other entries of $x$ to be zeros. $\|x\|_2=\sqrt{x^Tx}$ and $ \|x\|_\infty =\max_{1\leq i\leq n} |x_i|$ are  the $\ell_2$-norm and $\ell_\infty$-norm, respectively.

\subsection{Traditional OMP algorithm}\label{sub12}

Referred to as stagewise regression, the OMP algorithm first appeared in the area of statistics several decades ago.  It was introduced to signal processing community around 1993 (see, e.g.,  \cite{PRK93,DMZ94,MZ93}) and later to approximation theory (e.g., \cite{DMA97,D98,T03}). The OMP algorithm is stated as follows, which iteratively
selects a vector basis that is best correlated to
the residual at the current iterate.

\begin{algorithm}\textbf{OMP Algorithm.} Perform the following steps until a certain stopping criterion is satisfied:
\begin{itemize} \item[S1] Initialization:  $ x^{(0)} =0 $ and  $ S^{(0)} =\emptyset.$

\item[S2]  Given $x^{(p)} $ and  $ S^{(p)},$    let
$$ S^{(p+1)} = S^{(p)} \cup L_1(A^T (y-Ax^{(p)})),$$
$$  x^{(p+1)}= \mathop{\arg\min}_{z\in \mathbb{R}^n} \{\| y-Az\|_2:  ~ \textrm{supp}  (z) \subseteq S^{(p+1)} \}.
$$
\end{itemize}
\end{algorithm}

The advantage of OMP lies in  its simple structure allowing a fast
implementation at a low computational cost. OMP has stimulated the development of several other compressed sensing algorithms, including the   subspace pursuit (SP) \cite{DM09}, compressive sampling matching pursuit (CoSaMP) \cite{NT09}, multipath matching pursuit \cite{KWS14},  stagewise orthogonal matching pursuit (StOMP) \cite{DTDS12},  regularized orthogonal matching pursuit (ROMP) \cite{NV09, NV10}, general orthogonal matching pursuit (gOMP) \cite{WKS12}, weak orthogonal matching pursuit (WOMP) \cite{T00}, stagewise weak orthogonal matching pursuit (SWOMP) \cite{CXYC20} and constrained matching pursuit (CMP)\cite{SM20}. In each iteration, the OMP algorithm only select one index corresponding to the largest entry in magnitude of $ A^T (y-Az)$, the gradient of the error metric $   \|y-Az\|_2^2/2 .$  Thus the OMP algorithm does  not utilize the gradient information efficiently
 when the gradient vector possesses several large entries in magnitude whose absolute values are  close to each other. The gOMP algorithm is a direct generalization of OMP, allowing $N$ indices to be selected simultaneously in every iteration, where $1\leq N <k$ is a prescribed integer number. However, such a selection rule remains very rigid, and it might significantly increase the chance for a wrong index being selected.  For instance, if the gradient vector of error matric is $\tau$-compressible, where $ \tau < N,$  then the $N$ indices selected by gOMP would include an index corresponding to a vanished or insignificant entry of the gradient vector.
   The SP and CoSaMP in their nature are not close family members of the OMP-type methods. At each step, they iteratively and simultaneously search $k$ indices as the approximation to the support of signal. The StOMP, ROMP, WOMP and SWOMP adopt other different index selection rules to possibly enhance the performance of the standard OMP procedure. A common feature of these methods is that they allow the algorithm to select more than one indices in each iteration. WOMP is remarkably different from OMP in the sense that the index for the largest absolute entry of $ A^T (y-Az)$ might be excluded during its iterations. StOMP and SWOMP might select too many indices in a single iteration which admits a high risk for wrong indices to be selected.    Recently, the CMP method, as a modification of OMP, was proposed for solving more complicated  recovery problems with certain constraints. In this paper, we propose further modifications of OMP, which dynamically select a few indices among the $k$ largest components in magnitude of $ A^T (y-Az).$  We also allow the algorithms to adjust the selected index sets at every iteration by performing a shrinkage step when the total number of selected indices goes beyond the sparsity level of the signal (see the description of the proposed algorithm in section \ref{sect2}).

\subsection{Existing results for OMP}

The theoretical performance  of OMP-type methods can be analyzed in terms of coherence tools  (e.g.,  \cite{GMS03, T04, T06, TG07, E10}).   However, the   restricted isometry property (RIP) of the measurement matrix becomes relatively more popular than  coherence  for the  analysis of various compressed sensing algorithms including OMP. The RIP is defined as follows.

\begin{Def}\cite{CT05} \label{Def1.1} Let $A $ be an $m\times n$  matrix.  The  restricted isometry constant (RIC)  denoted by $\delta_q $ is the smallest number $\delta \geq 0$ such that
$$ (1-\delta) \|z\|^2_2 \leq \|Az\|^2_2
\leq (1+\delta)\|z\|^2_2 $$  for any $q$-sparse vector
$z\in \mathbb{R}^n. $ $A$ is said to satisfy the restricted isometry property (RIP) of order $q  $ if $ \delta_q<1.$
\end{Def}

It was shown in \cite{DW10} that OMP can identify the support of a $k$-sparse signal in exact $k$  iterations if  $ \delta_{k+1} < 1/(3\sqrt{k}).$  This condition was relaxed to   $ \delta_{k+1} < 1/(1+  \sqrt{2k}) $ in \cite{HZ11} and  $   \delta_{k+1} < 1/( \sqrt{k} +1 )$   in \cite{MS12} and \cite{WS12}.  This condition   was further improved   to  $   \delta_{k+1} < (\sqrt{4k+1}-1)/ (2k)   $ in \cite{CW14}, which is situated between $1/( \sqrt{k} +1 )$ and  $1/ \sqrt{k}.  $  Examples are given in \cite{MS12, WS12} to show that the OMP algorithm might fail to achieve the correct support identification in exact $k$ iterations when $\delta_{k+1} =  1/ \sqrt{k}.$   In noiseless situations (i.e., the measurements are accurate), it was   shown that  the condition $ \delta_{k+1} < 1/\sqrt{k+1}$ is sharp to  guarantee the success of OMP for $k$-sparse data reconstruction in exact $k$ iterations \cite{M15}. In noisy settings (the measurements are inaccurate), the guaranteed performance of OMP has been shown in \cite{CW11} and \cite{WHC13} under the  condition $ \label{CON01} \delta_{k+1} < 1/( \sqrt{k} +1 )$  and a certain assumption on the target signal.   Another improvement was achieved recently in \cite{WZLT17}, and it was shown that the condition $\delta_{k+1} < 1/  \sqrt{k  +1}   $ together with a condition on the signal is  sufficient for OMP to correctly identify the support of a $k$-sparse vector in exact $k$ iterations. Moreover, the stability of OMP was also studied under the RIP in \cite{Z11}.

 While CoSaMP and SP are also  motivated by OMP,  the two  algorithms equipped with a hard thresholding operator are very different from OMP. The interested readers can find the latest development of CoSaMP and SP algorithms in such reference as \cite{CW14b, FR13, SXL14, ZL20}, and can also find the results for other modifications of OMP such as StOMP, MMP, ROMP, gOMP and SWOMP. Here we only mention a few results for gOMP since it includes OMP as a special case and it is the closest kin to OMP.
The gOMP algorithm was introduced first in \cite{WKS12} (see also \cite{LT12, M11}).  It was shown in  \cite{WKS12} that  the condition $ \delta_{Nk} < \sqrt{N}/(\sqrt{k}+3\sqrt{N}) $ is sufficient for the gOMP algorithm to reconstruct the $k$-sparse signal  in $k$ iterations. This condition was relaxed in  \cite{LT12} and  \cite{SDC13}  and  was further relaxed  to $ \delta_{Nk} <  \sqrt{N}/(\sqrt{k}+1.27\sqrt{N}) $ in  \cite{SLPL14}. The reconstruction error via gOMP in noisy scenarios has  been investigated in \cite{LSWL15,W15,WZLT17}.

\subsection{Contribution of the paper}
 The main purpose of this paper is to propose an enhanced modification of OMP for sparse data reconstruction called \emph{dynamic orthogonal matching pursuit} (DOMP) and its  enhanced counterpart called \emph{enhanced dynamic orthogonal matching pursuit} (EDOMP).
  The proposed algorithms dynamically select vector bases  to reconstruct the target data according to the major gradient information of the error metric at every iteration. As a result, the gradient information is sufficiently and efficiently exploited to identify the support of the target data.  The reconstruction error bound via DOMP is established under the RIP assumption. This bound also implies the finite convergence of  DOMP in large-scale compressed sensing scenarios. Moreover, the numerical performance of the proposed algorithms is evaluated through random problem instances. The results indicate that due to the efficient usage of significant gradient information, the proposed algorithms are fast and robust for sparse data reconstruction and are very comparable to several state-of-art algorithms in this field.

The paper is organized as follows. The proposed algorithms are described in Section \ref{sect2}.   An approximation counterpart of the projection problem in the proposed algorithm is give in Section \ref{sect3}, which is used to facilitate the theoretical analysis of the algorithm.   Section \ref{sect4} is devoted to such a theoretical analysis which leads to the reconstruction error bound for the DOMP algorithm.  The finite convergence of  DOMP for  large-scale compressed sensing problems is also discussed is Section \ref{sect4}. Numerical results  are demonstrated in  Section \ref{sect5}.

\section{Dynamic orthogonal matching pursuit}\label{sect2}

 The OMP algorithm only select one index corresponding to the largest magnitude of the gradient of error metric at every iteration. To increase the chance for the correct support of the target data being identified, it would be  helpful to select indices at every iteration in a dynamic and adaptive manner. It is worth pointing out that this idea has been exploited in StOMP and SWOMP, but the total number of indices being selected at every iteration of these methods might increase too fast and exceed the sparsity level of the signal. This renders  the algorithm either requires extra effort to reconstruct the signal or fails to  reconstruct the signal.  Thus we propose the following \emph{dynamic orthogonal matching pursuit} (DOMP) method, which may  efficiently utilize  the underlying gradient information and control the number of  selected indices at each step so that it does not exceed the sparsity level of signal. When the cardinality of selected indices does exceed the sparsity level, we may further enhance the algorithm by including a shrinkage step to bring the cardinality down to the sparsity level.

\begin{algorithm}
 \textbf{DOMP Algorithm.} Perform the following steps until a certain stopping criterion is satisfied:
\begin{itemize} \item[S1] Initialization: Set $ x^{(0)} =0  $, $ S^{(0)} =\emptyset $ and $ r^{(0)}= A^T (y-A x^{(0)})=A^Ty.$  Choose parameter $ \gamma \in (0, 1]. $
\item[S2]  Given $x^{(p)}$, $ S^{(p)} $ and $ r^{(p)}= A^T(y-Ax^{(p)}),$ let
 $$ \Theta^{(p)} = \{ i:  ~ i \in L_{k} (r^{(p)} ), ~ |(r^{(p)})_i| \geq  \gamma \|r^{(p)} \|_\infty \},$$ and then set
\begin{align}   S^{(p+1)}   &  = S^{(p)} \cup \Theta^{(p)}  ,   \tag{DOMP1}   \\
    x^{(p+1)}  &  = \mathop{\arg\min}_{z\in \mathbb{R}^n} \{\| y-Az\|_2:  ~ \textrm{supp}  (z) \subseteq S^{(p+1)} \}.    \tag{DOMP2}
 \end{align}
\end{itemize}
\end{algorithm}

Two typical stopping criteria can be used in DOMP depending on the application environment: The algorithm  may stop in prescribed maximum number of iterations, and may also stop when $\|y-A x^{(p)}\|_2 \leq \varepsilon^* $  or  $ \| r^{(p)}\|_2 \leq \varepsilon^*,$  where $ \varepsilon^* >0$ is a prespecified tolerance.

Since $ x^{(p)}$  is an optimal solution to the convex optimization problem $$\min  \{\| y-Az\|_2:   \textrm{supp}  (z) \subseteq S^{(p)}  \},$$ by the first-order optimality condition, the iterates generated by DOMP satisfy that
\begin{equation} \label{OPT} (r^{(p)})_{S^{(p)}} = 0 ~ \textrm{ for any } p\geq 1 . \end{equation}
This property will be frequently used in later analysis of the algorithm.

The DOMP algorithm dynamically select the vector basis according to the criterion $ |(r^{(p)})_i| \geq \gamma\|r^{(p)} \|_\infty $ for $ i \in L_k(r^{(p)}).$ This criterion guarantees that  only the indices corresponding to the significant entries of $r^{(p)}$ can be used to identify the support of target data.   In general,  $ \Theta^{(p)} $ contains more than one elements corresponding to the indices of a few largest magnitudes of $ r^{(p)}.$ Clearly, even when $\gamma=1,$    the DOMP algorithm may not necessarily be the same as OMP since the set $\Theta^{(p)}$ may still contain more than one indices  if $ r^{(p)} $ possesses multiple entries whose absolute values are equal to $\|r^{(p)}\|_\infty . $  The cardinality of $ \Theta^{(p)}$  can be any value between 1 and $k,$ depending on the prescribed value of $\gamma$ and the number of significant entries of $ r^{(p)}.$  Such a dynamic selection of indices efficiently utilizes the gradient information of the error metric, and thus it might speed up the algorithm by increasing the chance for the correct  support of  target data being identified at every iteration.
After a few iterations are performed, the cardinality of $ S^{(p+1)}$ in DOMP might be larger than $k.$  In this case,  it is useful to perform certain shrinkage on the vector generated by (DOMP2) in order to maintain the $k$-sparsity of iterates so that they are feasible to the  problem (\ref{L0}). This consideration leads to the following \emph{enhanced dynamic orthogonal matching pursuit} (EDOMP), which is   different from existing modifications of OMP.

\begin{algorithm}
\textbf{EDOMP Algorithm.}  Perform the following steps until a certain stopping criterion is satisfied:
\begin{itemize} \item[S1] Initialization: Set $ x^{(0)} =0  $, $ S^{(0)} =\emptyset $, $ r^{(0)}=A^Ty $ and a parameter $ \gamma \in (0, 1]. $
\item[S2]  Given $x^{(p)}$, $ S^{(p)} $ and $ r^{(p)}= A^T(y-Ax^{(p)}),$ let
 $$ \Theta^{(p)} = \{ i: ~  i \in L_{k} (r^{(p)}), ~ |(r^{(p)})_i| \geq  \gamma \|r^{(p)}\|_\infty\}.$$ Set
$ S^{(p+1)} = S^{(p)} \cup \Theta^{(p)}  $ and
$$  \widetilde{x}^{(p+1)}= \mathop{\arg\min}_{z\in \mathbb{R}^n} \{\| y-Az\|_2:  ~ \textrm{supp}  (z) \subseteq S^{(p+1)} \}.    $$
If $|S^{(p+1)}|\leq k$, set $ x^{(p+1)}= \widetilde{x}^{(p+1)};$ otherwise, set $ Q=  L_k ( \widetilde{x}^{(p+1)})  $ and
$$ x^{(p+1)}= \textrm{arg}\min_{z\in \mathbb{R}^n} \{\| y-Az\|_2:  ~ \textrm{supp}  (z) \subseteq  Q \}. $$
\end{itemize}
\end{algorithm}

Numerical results in Section \ref{sect5} indicate that the capability of DOMP and EDOMP in sparse data reconstruction is generally stronger than OMP (see section \ref{sect5} for details). In the next section, we perform a theoretical analysis to establish a reconstruction error bound for the DOMP algorithm. Such an error bound measures the distance between the target data and the iterates generated by the algorithm. To facilitate this analysis, let us first define an  auxiliary optimization model.

\section{Approximation counterpart}\label{sect3}

Denote by $ S=L_k(x),$ and let $ x^{(p)},$ $ S^{(p)}, $ $  r^{(p)}, $ $  \Theta^{(p)}$  and $ S^{(p+1)}$   be defined in DOMP. Define
\begin{equation} \label{Lambda}  \Lambda^{(p)} := L_k (r^{(p)} ) \setminus (S^{(p)}\cup \Theta^{(p)} \cup S). \end{equation}
We now introduce an  approximation counterpart (AC) to the projection problem (DOMP2), which generates the index set $ \widehat{S}^{(p+1)} $ and the vector $ \widehat{x}^{(p+1)}.$

\begin{algorithm}
\textbf{Approximation counterpart of (DOMP2):}
Let  $x^{(p)}$, $ S^{(p)}$, $ \Theta^{(p)}$ and $ S^{(p+1)}$ be given in DOMP, and let $ \Lambda^{(p)}$ be defined by (\ref{Lambda}). Let
\begin{equation}  \widehat{S}^{(p+1)}    = S^{(p+1)} \cup \Lambda^{(p)},   \tag{AC1}  \end{equation}
 \begin{equation} \widehat{x}^{(p+1)}   = \mathop{\arg\min}_{z\in \mathbb{R}^n} \{ \| y-Az\|_2^2 + \sigma \|z_{\Lambda^{(p)}}\|_2^2: ~  \textrm{supp} (z) \subseteq \widehat{S}^{p+1} \},  \tag{AC2} \end{equation}
where   $ \| z_{\Lambda^{(p)}}\|_2^2 = \sum_{i\in \Lambda^{(p)}} z_i^2 $ and $ \sigma$ is a given large positive parameter.\
\end{algorithm}

When $r^{(p)} =0,$ $ x^{(p)}$ is already a global solution to the problem (\ref{L0}). Thus, to analyze the DOMP  algorithm, we assume without loss of generality that $ r^{(p)} \not=0 ,$ in which case we can see from the definition of $\Theta^{(p)} $ that  $ (r^{(p)})_i \not=0$ for any $ i\in \Theta^{(p)}. $  This together with (\ref{OPT}) implies that
\begin{equation}  \label{CSSS}    S^{(p)} \cap \Theta^{(p)} =\emptyset. \end{equation}
Since $ S^{(p+1)}= S^{(p)}\cup \Theta^{(p)}$, we also see from (\ref{Lambda}) that
\begin{equation} \label{CSSS12}  \Lambda^{(p)}\cap S^{(p+1)}  =\emptyset.   \end{equation}
 Thus
   \begin{equation} \label{DDTT}  |S^{(p+1)}|= |S^{(p)}| + |\Theta^{(p)}|,  ~|\widehat{S}^{(p+1)}|= |S^{(p+1)}| + |\Lambda^{(p)}|.
   \end{equation}

The solution of the optimization problem in  (AC2) depends on $ \sigma,$  and thus $ \widehat{x}^{(p+1)}$ should be written as $ \widehat{x}^{(p+1)} (\sigma) .$ For notational convenience, however,  we simply use $ \widehat{x}^{(p+1)}$ to denote the solution of (AC2). The above-described approximation counterpart is only used as an auxiliary tool to support the analysis of DOMP in the next section.
From  (AC2), one can see that for a sufficiently large  $ \sigma,$ the components of $ \widehat{x}^{(p+1)}$ supported on $ \Lambda^{(p)}$ would  vanish completely or be sufficiently small. Thus $\widehat{x}^{(p+1)}$  can be made arbitrarily close to $ x^{(p+1)}$ (the iterate generated by DOMP) provided that  $ \sigma $ is sufficiently large. The following lemma makes this rigorous.

 \begin{Lem} \label{proximity}
 At the iterate $ x^{(p)}$, let $ x^{(p+1)} $ and $  S^{(p+1)}$ be generated by DOMP, and let  $   \widehat{S}^{(p+1)} $ and $ \widehat{x}^{(p+1)} $be defined in (AC1) and (AC2) accordingly. Then one has
   \begin{equation} \label{AEST2} \|A (\widehat{x}^{(p+1)} -x^{(p+1)})\|_2  \leq    C \vartheta_{(y,A)}  ,
\end{equation}
 where  $ C := \sqrt{\frac{2\| A\|_2} {\sqrt{\sigma}}  + \frac{\|A\|_2^2} { \sigma }    }  + \frac{\|A\|_2}{\sqrt{\sigma}}$  and $  \vartheta_{(y,A)}  $ is the constant  defined as
 \begin{equation} \label{DN}   \vartheta_{(y,A)}:  =  \max_{ U\subseteq \{1,\cdots, n\},
 U \not= \emptyset  } (\min_{z\in \mathbb{R}^n} \{ \| y-Az\|_2:   ~ \textrm{supp} (z) \subseteq U \} ).     \end{equation}

\end{Lem}

\emph{Proof.} Suppose that  $ x^{(p)},  \Theta^{(p)}$ and $   x^{(p+1)}$  are  generated by DOMP. Let $ \Lambda^{(p)}$ be defined by (\ref{Lambda}) and $ \widehat{x}^{(p+1)}$ be a solution to the minimization problem (AC2). Note that  $x^{(p+1)}$ is a feasible solution to the minimization problem   (AC2), by the optimality of $ \widehat{x}^{(p+1)},$ we have
\begin{align} \label{EST}  \|  y   &  -     A \widehat{x}^{(p+1)}\|_2^2 + \sigma \|(\widehat{x}^{(p+1)})_{\Lambda^{(p)}} \|_2^2
\nonumber \\ &  \leq \| y-A x^{(p+1)}\|_2^2  + \sigma \|(x^{(p+1)})_{\Lambda^{(p)}} \|_2^2  \nonumber \\
& =  \| y-A x^{(p+1)}\|_2^2 , \end{align} where   the  equality  follows from  (\ref{CSSS12}) which implies that $(x^{(p+1)})_{\Lambda^{(p)}} =0.$ Thus
\begin{align*}   \sigma \|(\widehat{x}^{(p+1)})_{\Lambda^{(p)}} \|_2^2   &   \leq \| y-A x^{(p+1)}\|_2^2 \\
  &  = \min_{z\in \mathbb{R}^n} \{ \| y-Az\|_2^2:   \textrm{supp} (z) \subseteq  S^{(p+1)} \} \\
  &  \leq  \vartheta_{(y,A)}^2,
\end{align*}
where $ \vartheta_{(y,A)} $ is the constant defined in (\ref{DN}).  The inequalities above imply that
\begin{equation} \label{B2}  \|(\widehat{x}^{(p+1)})_{\Lambda^{(p)}} \|_2 \leq   \frac{ \vartheta_{(y,A)}} {\sqrt{\sigma}}, ~\| y-A x^{(p+1)}\|_2 \leq  \vartheta_{(y,A)}.  \end{equation}
We now bound $ \|A\left(x^{(p+1)} - \widehat{x}^{(p+1)}\right)\|_2. $
From  (\ref{EST}), we see that
\begin{equation} \label{EST2}   \| y-A \widehat{x}^{(p+1)}\|_2   \leq \| y-A x^{(p+1)}\|_2 .  \end{equation}
 By (AC1) and (\ref{CSSS12}),  we see that  $ \widehat{x}^{(p+1)}   = (\widehat{x}^{(p+1)})_{\widehat{S}^{(p+1)}}   = (\widehat{x}^{(p+1)})_{S^{(p+1)}} + (\widehat{x}^{(p+1)})_{\Lambda^{(p)}}. $  Thus by the triangular inequality, we have
\begin{align} \label{DIN}    \| y   &    -A \widehat{x}^{(p+1)}\|_2  \nonumber  \\
&  = \| y-A [(\widehat{x}^{(p+1)})_{S^{(p+1)}} + (\widehat{x}^{(p+1)})_{\Lambda^{(p)} }]\|_2 \nonumber  \\
 &  \geq  \| y-A (\widehat{x}^{(p+1)})_{S^{(p+1)}}\|_2  - \| A(\widehat{x}^{(p+1)})_{\Lambda^{(p)} }\|_2.
\end{align}
Merging  (\ref{EST2}) and (\ref{DIN}) leads to
$$ \| y-A (\widehat{x}^{(p+1)})_{S^{(p+1)}}\|_2 \leq \| y-A x^{(p+1)}\|_2 + \| A(\widehat{x}^{p+1})_{\Lambda^{(p)}}\|_2, $$
which together with (\ref{B2}) implies that
 \begin{align}   \label{CC1}
 &    \|y     -A (\widehat{x}^{(p+1)})_{S^{(p+1)}}\|_2^2 \nonumber \\
 & \leq  (\| y-A x^{(p+1)}\|_2 +  \| A(\widehat{x}^{(p+1)})_{\Lambda^{(p)}}\|_2 )^2  \nonumber \\
    & = \| y-A x^{(p+1)}\|_2^2 \nonumber \\
     &    ~~ + 2\| y-A x^{(p+1)}\|_2 \| A(\widehat{x}^{(p+1)})_{\Lambda^{(p)}}\|_2     + \| A(\widehat{x}^{(p+1)})_{\Lambda^{(p)}}\|_2^2  \nonumber\\
    & \leq \| y-A x^{(p+1)}\|_2^2  \nonumber \\
    & ~~ + 2 \vartheta_{(y,A)} \| A\|_2 \|(\widehat{x}^{(p+1)})_{\Lambda^{(p)}}\|_2
      + \| A\|_2^2 \|(\widehat{x}^{(p+1)})_{\Lambda^{(p)}}\|_2^2  \nonumber \\
     & \leq \| y-A x^{(p+1)}\|_2^2  \nonumber \\
      & ~~ + 2 \vartheta_{(y,A)} \| A\|_2 \left(\frac{\vartheta_{(y,A)}}{\sqrt{\sigma}} \right)
     + \| A\|_2^2 \left(\frac{\vartheta_{(y,A)}}{\sqrt{\sigma}}\right)^2 \nonumber \\
    & = \| y-A x^{(p+1)}\|_2^2 + \left( \frac{2 \| A\|_2} {\sqrt{\sigma}}  + \frac{\|A\|_2^2} { \sigma }  \right) (\vartheta_{(y,A)})^2.
   \end{align}
By Taylor's expansion, we also have that
\begin{align}  \label{CC2}  &  \| y     -A (\widehat{x}^{(p+1)})_{S^{(p+1)}}\|_2^2 = \| y-A x^{(p+1)}\|_2^2 \nonumber \\
&    + 2[-A^T (y-Ax^{(p+1)})]^T [(\widehat{x}^{(p+1)})_{S^{(p+1)}} -x^{(p+1)}]  \nonumber\\
 &  + [(\widehat{x}^{(p+1)})_{S^{(p+1)}}-x^{(p+1)}]^TA^TA [(\widehat{x}^{(p+1)})_{S^{(p+1)}}-x^{(p+1)}] \nonumber \\
& =\| y-A x^{(p+1)}\|_2^2   +  \|A [(\widehat{x}^{(p+1)})_{S^{(p+1)}}-x^{(p+1)}]\|_2^2,
\end{align}
where the last equality follows from (\ref{OPT}) and the fact that $ \textrm{supp} [(\widehat{x}^{(p+1)})_{S^{(p+1)}}$ $ - x^{(p+1)}] $ $ \subseteq S^{(p+1)} . $
Combining (\ref{CC1}) and (\ref{CC2}) yields
\begin{equation} \label{DDDRRR}  \|A [(\widehat{x}^{(p+1)})_{S^{(p+1)}}-x^{(p+1)}]\|_2 ^2  \leq  \left[ \frac{2\| A\|_2} {\sqrt{\sigma}}  + \frac{\|A\|_2^2} { \sigma }   \right]  (\vartheta_{(y,A)})^2  . \end{equation}
By (AC1) and (\ref{CSSS12}), i.e., $ \widehat{S}^{(p+1)} = S^{(p+1)} \cup \Lambda^{(p)} $ and $   S^{(p+1)} \cap \Lambda^{(p)} =\emptyset, $ we immediately see that $\left(\overline{ \widehat{S}^{(p+1)} }\right) \cap  \Lambda^{(p)} =\emptyset  $ and
$  \overline{S^{(p+1)}} =\overline{ \widehat{S}^{(p+1)}   \setminus \Lambda^{(p)} } = \overline{ \widehat{S}^{(p+1)} } \cup  \Lambda^{(p)},  $
and hence
\begin{align*} (\widehat{x}^{(p+1)})_{\overline{S^{(p+1)}}}    = (\widehat{x}^{(p+1)})_{\overline{ \widehat{S}^{(p+1)} }} +    (\widehat{x}^{(p+1)})_{\Lambda^{(p)}}
    = (\widehat{x}^{(p+1)})_{\Lambda^{(p)}},
\end{align*}  where the last equality follows from the fact that  $(\widehat{x}^{(p+1)})_{\overline{ \widehat{S}^{(p+1)} }}  =0 . $
This together with (\ref{B2}), (\ref{DDDRRR}) and $ ( x^{(p+1)})_{\overline{S^{(p+1)}}}=0 $  implies that
 \begin{align*}  & \|A (\widehat{x}^{(p+1)} -x^{(p+1)})\|_2 \\
 & =  \|A [(\widehat{x}^{(p+1)} -x^{(p+1)})_{S^{(p+1)}} +  (\widehat{x}^{(p+1)} -x^{(p+1)})_{\overline{S^{(p+1)}}} ]\|_2 \\
& = \|A [(\widehat{x}^{(p+1)})_{S^{(p+1)}} -x^{(p+1)}] +  A(\widehat{x}^{(p+1)})_{\Lambda^{(p)}} \|_2\\
& \leq \|A [(\widehat{x}^{(p+1)})_{S^{(p+1)}} -x^{(p+1)}] \|_2 +\| A(\widehat{x}^{(p+1)})_{\Lambda^{(p)}} \|_2 \\
& \leq \vartheta_{(y,A)} \sqrt{ \frac{2\| A\|_2 } {\sqrt{\sigma}}  + \frac{\|A\|_2^2} { \sigma }  } + \|A\|_2 \frac{\vartheta_{(y,A)}}{\sqrt{\sigma}}
\end{align*}
which is the inequality (\ref{AEST2}).\hfill $\Box $

A basic property of the optimization problem (AC2) is given in the next lemma which follows directly from the optimality condition of convex optimization. The proof of the lemma is omitted.

\begin{Lem} \label{optimality}
Let $ \widehat{x}^{(p+1)}$ be a minimizer of (AC2), i.e., $$\widehat{x}^{(p+1)}= \mathop{\arg\min}_{z\in \mathbb{R}^n} \{ \| y-Az\|_2^2 + \sigma \|z_{\Lambda^{(p)}}\|_2^2 : ~ \textrm{supp}  (z) \subseteq \widehat{S}^{(p+1)} \}, $$
where  $ \widehat{S}^{(p+1)}   $ is defined by (AC1). Then
\begin{equation} \label{EQ0001} \left[-  A^T(y-A \widehat{x}^{(p+1)})  + \sigma ( \widehat{x}^{(p+1)} )_{\Lambda^{(p)}} \right]_{ \widehat{S}^{(p+1)} }   =0. \end{equation}
\end{Lem}

\section{Reconstruction error bounds}\label{sect4}

 In this section, we establish the reconstruction error bound for DOMP. The efficiency of the algorithm for sparse data reconstruction can be interpreted via such an error bound which measures the distance of the generated solution and target data. The error bound may  also  imply the stability and convergence of the algorithm under certain conditions.
Let us first recall some properties of $\delta_t$ that will be used to establish the desired error bound.

\begin{Lem} {\rm \cite{FR13} }\label{Lem-Basic} Let $  u \in \mathbb{R}^n $ and   $\Lambda \subseteq \{1,2, \dots, n\}.$ Then the following three statements are true:
\begin{itemize}
\item[{\rm (i)}]  If $|\Lambda \cup  {\rm supp}  (u) | \leq t,$ then
$ \| [(I-A^TA)u]_{\Lambda}\|_2  \leq \delta_t \|u\|_2 .
$

\item[{\rm (ii)}]   If $ \Lambda \cap {\rm supp} (u) = \emptyset$ and  $ | \Lambda \cup  {\rm supp} (u) | \leq t,$ then $   \|(A^T A u)_\Lambda\|_2\leq \delta_t \| u\|_2.  $

\item[{\rm (iii)}]  If $|\Lambda |\leq t,$ one has $\|(A^T u)_{\Lambda}\|_2 \leq \sqrt{1+\delta_t} \|u\|_2. $
\end{itemize}
\end{Lem}

It should be pointed out that $ \delta_t$ is not the only tool that can be used to analyze algorithms. Other tools such as the spark \cite{E10}, mutual coherence \cite{E10},  null space property \cite{FR13} and range space property of $A^T$ \cite{Z18, Z13} might be also used to achieve this goal. However, we only use $\delta_t$ to establish the main result in this paper.

\subsection{Main result for DOMP}

The purpose of this section is to answer the question:  How does the reconstruction error decay in the course of iterations?  In other words, we would like to measure the quality of the iterate, generated by the algorithm, as an approximation to the target data.
The result for DOMP is summarized in Theorem \ref{Thm-DOMP} below,  whose proof is postponed to the end of the next subsection after we establish some helpful technical results.

\begin{Thm} \label{Thm-DOMP} Suppose that  $y: =Ax+\nu$ are the measurements of the target data $x
\in \mathbb{R}^n$ where $ \nu $ are the measurement errors. Let  $ \gamma \in(0,1] $ be a  given constant.    If the RIC  of the measurement matrix $A$  satisfies
\begin{equation} \label{RIPNewB} \delta_{ck}  <   \frac{1} {\sqrt{ 1+ \left(1+ \sqrt{ 1+k\gamma^2}\right )^2  }} ,  \end{equation}
where $ c> 2 $ is an integer number, then at the $p$th iteration, provided that $ |S^{p+1}| \leq  (c-2)k,$  the iterate $x^{(p+1)}$ generated by DOMP approximates $x_S$ with error
\begin{align}\label{FERR} \|x^{(p+1)}- x_S\|_2 \leq  & \beta ^p \left((\varrho/\beta)^{k} \|x_S-x^{(0)}\|_2 \right)   + \tau \|\nu'\|_2  , \end{align}
where $ \nu' = Ax_{\overline{S}}+ \nu   $ and $ S= L_k(x). $ The constants $ \beta $ and $ \varrho$ are given respectively as
$$ \beta =  \frac{ \left( 1+\sqrt{1+ k\gamma^2} \right) \delta_{ck}   }  {\sqrt{1- \delta^2_{ck} } } <1,  ~ \varrho = \sqrt{ \frac{1+\delta_{ck} } { 1-\delta_{ck}  }  }, $$
  and  $\tau  $ is a constant dependent only of $\delta_{ck}, $  $ k$ and $ \gamma .$
\end{Thm}

\begin{Rem} If $ x$ is $k$-compressible with very small tail $ \|x_{\overline{S}}\|_2$ and the measurements are accurate enough, then  $ \|\nu'\|_2$ would be very small. In particular, if $ x$ is $k$-sparse and the measurements are accurate, then $ \| \nu'\|_2=0.$  From (\ref{FERR}), the reconstruction quality in these cases would completely depend on the iterations being executed.  We also note that $ |S^{(p+1)}| $ is the total number of indices being selected after $p$ iterations. At the $p$th iteration, the indices in $ \Theta^{(p)}$ are added to $ S^{(p)}$ to form the set   $ S^{(p+1)}.$ By (\ref{CSSS}),   $ |S^{(p+1)}|= \sum_{i=1}^p  |\Theta^{(p)} |  \leq pk,$  thus the largest integer number $ p^*$ satisfying $|S^{(p^*+1)}|\leq (c-2)k $   would be at least $ (c-2).$ This means when $ c \gg 2$ the error bound (\ref{FERR}) is satisfied for a large number $ p =p^*$ and thus $\|x^{(p)}-x_S\|_2  \approx 0.$ Thus $ x^{(p)}$ is a quality reconstruction of $ x_S  $ after the algorithm is performed enough iterations.
\end{Rem}

 The lemma below follows immediately from Theorem \ref{Thm-DOMP}.

\begin{Cor} Let $y:=Ax+\nu$ be the measurements of the target data $x\in \mathbb{R}^n $ where $ \nu $ are the measurement errors. Let  $ \gamma \in(0,1] $ be a  given constant. If the matrix $A$ satisfies the RIP of order $ t^* > 2k$ where  \begin{equation} \label{RIP-New-C}  t^* := \max\left\{t: \delta_t <  \frac{1} { \sqrt{ 1+ \left(1+\sqrt{1+k\gamma^2} \right)^2   }} \right\},  \end{equation}
then at the $p$th iteration, provided $ |S^{p+1}| \leq  t^*-2k,$  the iterate $x^{(p+1)}$ generated by DOMP approximates $x_S$ with error
\begin{align*} \|x^{(p+1)}- x_S\|_2 \leq  &  \hat{\beta} ^p \left((\hat{\varrho}/\hat{\beta})^{k} \|x_S-x^{(0)}\|_2 \right) + \hat{\tau} \|\nu'\|_2,
\end{align*}
where $ \nu ' = A x_{\overline{S}}+\nu  $  and $ S= L_k(x). $   The constants $ \hat{\beta} $ and  $ \hat{\varrho}$  are given respectively as
$$ \hat{\beta} =  \frac{ \left( 1+\sqrt{1+ k\gamma^2} \right) \delta_{t^*}   }  {\sqrt{1- \delta^2_{t^*} } }<1,  ~ \hat{\varrho} = \sqrt{ \frac{1+\delta_{t^*} } { 1-\delta_{t^*}  }  }, ~$$
  and   $\hat{\tau}$  is a  constant dependent only of $\delta_{t^*}$, $  k$ and $ \gamma. $
\end{Cor}

\vskip 0.15in

 The results above indicate that the quality of the iterate as an approximation to $ x_S$ can be measured provided that the total number of selected indices   is  lower than $t^*-2k.$     The RIC bounds (\ref {RIPNewB}) and (\ref{RIP-New-C}) depends on the parameter $ \gamma$. In theory,   a small  parameter $ \gamma$ can alleviate the requirement on measurement matrices since the smaller the parameter $ \gamma,$ the larger the right-hand sides of (\ref{RIPNewB}) and (\ref{RIP-New-C}).

\subsection{Proof of the main result for DOMP}

We first establish several useful technical results. The first one below is partially shown in \cite{ZL20}.

\begin{Lem}  \label{Lem-3A}
Given  constants $ \alpha_1, \alpha_2, \alpha_3\geq 0$ where $ \alpha_1<1,$ if $t$ satisfies that
\begin{equation} \label{Con1} t \leq \alpha_1 \sqrt{t^2+\alpha_2^2}+ \alpha_3, \end{equation}  then
  \begin{equation} \label{Con1a} t\leq  \frac{\alpha_1}{\sqrt{1-\alpha_1^2}}  \alpha_2 +  \frac{\alpha_3}{1-\alpha_1},
  ~ \sqrt{ t^2+ \alpha_2^2}  \leq \frac{\alpha_2}{\sqrt{1-\alpha_1^2}} +\frac{\alpha_3}{1-\alpha_1}.
   \end{equation}
\end{Lem}

\emph{Proof}. The first inequality in (\ref{Con1a}) was shown in \cite{ZL20} (see Lemma 4.1 therein). Thus we only need to show the second inequality in (\ref{Con1a}), which follows from the first inequality  and the fact that $ \sqrt{(a+c)^2 + b^2} \leq \sqrt{a^2+b^2 } +c $  for any $ a,b,c \geq 0. $ \hfill $ \Box $

 Some fundamental properties of DOMP  and the optimization problem (AC2) are given in the next two lemmas.

\begin{Lem} \label{Lem-ERR-01} Let $ y:= Ax+\nu , $   $ S^{(p+1)} $ be generated by DOMP,  $ \widehat{S}^{(p+1)} $  be defined in (AC1), and let  $ \widehat{x}^{(p+1)}$ be a minimizer of (AC2).  If
  $A$ satisfies that
$ \delta_{k+|\widehat{S}^{(p+1)}|} <1, $  then
\begin{align}\label{LEEE}  \|x_S  & -\widehat{x}^{(p+1)} \|_2  \nonumber \\
 & \leq \frac{ \|(x_S -\widehat{x}^{(p+1)})_{\overline{\widehat{S}^{(p+1)}}}  \|_2  }  {\sqrt{1- \delta^2_{k+|\widehat{S}^{(p+1)}| } } }
   + \frac{\sqrt{1+\delta_{|\widehat{S}^{(p+1)}|}}}{ 1-\delta_{k+|\widehat{S}^{(p+1)}| } }   \| \nu'   \|_2,
    \end{align}
    where $ \nu'= Ax_{\overline{S}}+ \nu  $ and $ S=L_k(x). $
\end{Lem}

\emph{Proof.}  By Lemma \ref{optimality},  we have that
$ [-A^T(y-A \widehat{x}^{(p+1)})+\sigma ( \widehat{x}^{(p+1)})_{\Lambda^{(p)}} ]_{\widehat{S}^{(p+1)}} =0.  $
 As $ y=Ax +\nu= A x_S +\nu' $ where $ \nu'= Ax_{\overline{S}}+ \nu, $ the equality above can be written as
$$ [-A^TA  (x_S-\widehat{x}^{(p+1)}) -A^T \nu' + \sigma ( \widehat{x}^{(p+1)})_{\Lambda^{(p)}}]_{\widehat{S}^{(p+1)}} =0. $$
 Note that $ [(\widehat{x}^{(p+1)} )_{\Lambda^{(p)}}]_{\widehat{S}^{(p+1)}} =  (\widehat{x}^{(p+1)})_{\Lambda^{(p)}}$ since $\Lambda^{(p)} \subseteq \widehat{S}^{(p+1)}.$
The equality above is equivalent to
\begin{align} \label{EEE200}   (x_S     -\widehat{x}^{(p+1)}     )_{\widehat{S}^{(p+1)}}  & = [(I-A^TA ) (x_S -\widehat{x}^{(p+1)})]_{\widehat{S}^{(p+1)}} \nonumber \\
  & ~~~  -(A^T \nu')_{\widehat{S}^{(p+1)}} +\sigma ( \widehat{x}^{(p+1)})_{\Lambda^{(p)}}.
\end{align}
By the definition of $\Lambda^{(p)}$ in (\ref{Lambda}), we see that $ S\cap \Lambda^{(p)} =\emptyset.$ Thus  $  (x_S)_{\Lambda^{(p)}} =0  $ and
\begin{align} \label{SSS200}
  [(x_S  & -\widehat{x}^{(p+1)}   )_{\widehat{S}^{(p+1)}}]^T   (\widehat{x}^{(p+1)})_{\Lambda^{(p)}} \nonumber \\
  & = [(x_S-\widehat{x}^{(p+1)})_{ \Lambda^{(p)} }]^T   (\widehat{x}^{(p+1)})_{\Lambda^{(p)}} \nonumber \\
  & = - \|(\widehat{x}^{(p+1)})_{\Lambda^{(p)}}\|_2^2 \leq 0.
\end{align}
Multiplying (\ref{EEE200}) by $(x_S-\widehat{x}^{(p+1)})_{\widehat{S}^{(p+1)}}$, using (\ref{SSS200}) and
Lemma \ref{Lem-Basic} (i) and (iii), we obtain
\begin{align} \label{SSS01}  &  \|(x_S      -\widehat{x}^{(p+1)})_{\widehat{S}^{(p+1)}}\|_2^2  \nonumber \\
 &  =   [(x_S-\widehat{x}^{(p+1)})_{\widehat{S}^{(p+1)}}]^T  [(I-A^TA )  (x_S-\widehat{x}^{(p+1)})]_{\widehat{S}^{(p+1)}}   \nonumber \\
 & ~~~ - [(x_S-\widehat{x}^{(p+1)})_{\widehat{S}^{(p+1)}}]^T(A^T \nu')_{\widehat{S}^{(p+1)}}  \nonumber\\
 & ~~~ + \sigma [(x_S-\widehat{x}^{(p+1)})_{\widehat{S}^{(p+1)}}]^T    (\widehat{x}^{(p+1)})_{\Lambda^{(p)}}   \nonumber\\
 & \leq [(x_S-\widehat{x}^{(p+1)})_{\widehat{S}^{(p+1)}}]^T  [(I-A^TA )  (x_S-\widehat{x}^{(p+1)})]_{\widehat{S}^{(p+1)}}   \nonumber \\
  & ~~~ - [(x_S-\widehat{x}^{(p+1)})_{\widehat{S}^{(p+1)}}]^T(A^T \nu')_{\widehat{S}^{(p+1)}}\nonumber\\
& \leq   \delta_{k+|\widehat{S}^{(p+1)}|} \|x_S-\widehat{x}^{(p+1)}\|_2 \|(x_S-\widehat{x}^{(p+1)})_{\widehat{S}^{(p+1)}}\|_2  \nonumber \\
& ~~~ +   \sqrt{1+\delta_{|\widehat{S}^{(p+1)}|}}\| \nu' \|_2 \| (x_S-\widehat{x}^{(p+1)})_{\widehat{S}^{(p+1)}} \|_2  \nonumber\\
 & =  \|(x_S-\widehat{x}^{(p+1)})_{\widehat{S}^{(p+1)}}\|_2 ( \delta_{k+|\widehat{S}^{(p+1)}|}  \|x_S-\widehat{x}^{(p+1)}\|_2 \nonumber\\
  & ~~~  +     \sqrt{1+\delta_{|\widehat{S}^{(p+1)}|}} \| \nu '\|_2).
\end{align}
If $ \|(x_S-\widehat{x}^{(p+1)})_{\widehat{S}^{(p+1)}}\|_2=0,$ then $\|x_S-\widehat{x}^{(p+1)}\|_2 = \|(x_S-\widehat{x}^{(p+1)})_{\overline{\widehat{S}^{(p+1)}}}\|_2,$ and hence the inequality (\ref{LEEE}) is satisfied trivially. Thus we assume without loss of generality that $ \|(x_S-\widehat{x}^{(p+1)})_{\widehat{S}^{(p+1)}}\|_2\not =0 .$ We cancel this term from (\ref{SSS01}) to obtain that
{\small \begin{align*}  &  \|(x_S     -\widehat{x}^{(p+1)})_{\widehat{S}^{(p+1)}}\|_2 \nonumber\\
 & \leq \delta_{k+|\widehat{S}^{(p+1)}|}   \|x_S-\widehat{x}^{(p+1)}\|_2 +   \sqrt{1+\delta_{|\widehat{S}^{(p+1)}|}}  \| \nu '\|_2 \nonumber   \\
 &= \delta_{k+|\widehat{S}^{(p+1)}|} \sqrt{\|(x-\widehat{x}^{(p+1)})_{\widehat{S}^{(p+1)}}  \|_2^2
     +  \|(x-\widehat{x}^{(p+1)})_{\overline{\widehat{S}^{(p+1)}}}  \|_2^2 }  \\
      & ~~~ +  \sqrt{1+\delta_{|\widehat{S}^{(p+1)}|}}  \|\nu' \|_2,
\end{align*} }
which is in the form of (\ref{Con1}) with $t= \|(x_S -\widehat{x}^{p+1})_{\widehat{S}^{(p+1)}}\|_2, $   $ \alpha_1 =\delta_{k+|\widehat{S}^{(p+1)}|}, ~ \alpha_2= \|(x_S -\widehat{x}^{(p+1)})_{\overline{\widehat{S}^{(p+1)}}}\|_2,  ~ \alpha_3 = \sqrt{1+\delta_{|\widehat{S}^{(p+1)}|}} \|\nu' \|_2 ,$
and \begin{align*} \|x_S  &  -\widehat{x}^{(p+1)}  \|_2 \\
  & = \sqrt{\|(x_S -\widehat{x}^{(p+1)})_{\widehat{S}^{(p+1)}}\|_2^2 + \|(x_S -\widehat{x}^{(p+1)})_{\overline{\widehat{S}^{(p+1)}}}\|_2^2 } \\
   &  = \sqrt{t^2+ \alpha_2^2} .
   \end{align*}
Thus it immediately follows from Lemma  \ref{Lem-3A} that
\begin{align*}    \| x_S -\widehat{x}^{(p+1)}  \|_2
    & \leq \frac{1}{\sqrt{1- \delta^2_{k+|\widehat{S}^{(p+1)}|}}} \|(x_S -\widehat{x}^{(p+1)})_{\overline{\widehat{S}^{(p+1)}}} \|_2 \\
    &  + \frac{ \sqrt{1+\delta_{|\widehat{S}^{(p+1)}|}} }{1- \delta_{k+|\widehat{S}^{(p+1)}|} } \| \nu' \|_2,
  \end{align*}
which is exactly the relation (\ref{LEEE}).     \hfill  $ \Box $

 Under the condition $ \Theta^{(p)} \cap S =\emptyset,$ we now  bound  the term $ \|(x_S- \widehat{x}^{(p+1)})_{ \overline{\widehat{S}^{(p+1)}} } \|_2 $ which is equal to $ \|(x_S- x^{(p)})_{ \overline{\widehat{S}^{(p+1)}} } \|_2  $ since $(\widehat{x}^{(p+1)})_{ \overline{\widehat{S}^{(p+1)}} }= 0 $ and $ (x^{(p)})_{ \overline{\widehat{S}^{(p+1)}} }=0 .$

\begin{Lem} \label{LEM-Final}
Let $ y: = Ax+\nu, $  and let  $ x^{(p)},$ $ S^{(p)},$ $  \Theta^{(p)}$ and $ S^{(p+1)}$ be generated by DOMP.    Let $\widehat{S}^{(p+1)}   $  be defined by (AC1), i.e.,  $\widehat{S}^{(p+1)}   =   S^{(p+1)}\cup  \Lambda^{(p)}  $ where $ \Lambda^{(p)}$ is defined by (\ref{Lambda}).  Suppose that  $ \Theta^{(p)} \cap S =\emptyset $ where $ S= L_k(x). $ If the matrix $A$ satisfies that $ \delta_{k+|\widehat{S}^{(p+1)}|} <1,$ then one has
\begin{align}\label{EELL02}  \| (x_S  -x^{(p)})_{\overline{ \widehat{S}^{(p+1)}}} \|_2   \leq  &  \rho^*(\delta_{2k+| S^{(p)}|}  \|x_S- x^{(p)}\|_2  \nonumber \\
     &  +  \sqrt{1+\delta_k}\|\nu'\|_2),
 \end{align}
where $ \rho^* =1+ \sqrt{1+ \gamma^2k},  $ $  \nu'= A x_{\overline{S}}+\nu  $    and $ \gamma \in (0, 1] $ is a given parameter in DOMP.
\end{Lem}

{\it Proof. } Let $ x^{(p)}$ be the current iterate and $ r^{(p)}, S^{(p)}, \Theta^{(p)}, S^{(p+1)}$ be defined in DOMP. Suppose that  $\Theta^{(p)}$ and $S$ are disjoint, i.e.,  \begin{equation} \label{COND0001} S\cap \Theta^{(p)} =\emptyset. \end{equation}
We denote by $ \widetilde{L}: = L_k(r^{(p)}) $ for notational convenience.
   Note that $|\widetilde{L} | =k =|S| .$   Thus \begin{align} \label{DDD}  |S \backslash \widetilde{L} |  = |S|- | S  \cap \widetilde{L} |
   = |\widetilde{L} |-| S  \cap \widetilde{L}|
   =|\widetilde{L} \backslash  S |. \end{align}
 This means the number of elements in $ S \backslash \widetilde{L} $ is equal to the number of elements in $ \widetilde{L} \backslash S .$
 By the definition of $\widetilde{L}$, the entries of $ r^{(p)} = A^T (y-Ax^{(p)})$ supported on $ \widetilde{L}   \backslash S $ are  among the $k$ largest magnitudes of $ r^{(p)}.$ This together with (\ref{DDD}) implies that
 \begin{equation} \label{ZZ01}  \|(r^{(p)})_{ S \backslash \widetilde{L} } \|_2  \leq  \|(r^{(p)})_{\widetilde{L}  \backslash  S } \|_2. \end{equation}
By (\ref{Lambda}), $  S \cap \Lambda^{(p)} =\emptyset $ which together with (\ref{COND0001}) implies  that
 \begin{equation} \label{ZZEE01}  S \cap \widehat{S}^{(p+1)} = S \cap (S^{(p)} \cup \Theta^{(p)} \cup \Lambda^{(p)}) = S \cap S^{(p)}.
 \end{equation}
 Note that  $ S = (S \backslash T ) \cup (S\cap T)$ for any set $T, $ where $ S  \setminus T $ and $ S\cap T$ are disjoint.   By setting $ T = \widehat{S}^{(p+1)}$ and $ T= \widetilde{L} $ respectively, we immediately obtain the  two relations:
  \begin{align}    \|(r^{(p)})_S\|_2^2    & =  \|(r^{(p)})_{ S \backslash \widehat{S}^{(p+1)}}\|_2^2
     + \|(r^{(p)})_{ S \cap \widehat{S}^{(p+1)}}\|_2^2,    \label{ZZEE02}  \\
      \|(r^{(p)})_S\|_2^2   &  =  \|(r^{(p)})_{ S \backslash \widetilde{L}}\|_2^2
    + \|(r^{(p)})_{ S \cap \widetilde{L}}\|_2^2. \label{ZZEE03}
  \end{align}
Note that $ x^{(p)}$ is a minimizer of the convex optimization problem
  $ \min\{\|y-Az\|_2: \textrm{supp} (z) \subseteq S^{(p)}\}. $ Thus $ x^{(p)}$ and $ S^{(p)}$ satisfy (\ref{OPT}), i.e.,
  \begin{equation} \label{OP-01}   (r^{(p)})_{S^{(p)}} = 0. \end{equation}
  This together with (\ref{ZZEE01})  implies that
 $ \|(r^{(p)})_{ S \cap \widehat{S}^{(p+1)}}\|_2^2=\|(r^{(p)})_{ S \cap S^{(p)}}\|_2^2 =0. $
 Thus, merging   (\ref{ZZEE02}) and (\ref{ZZEE03}) yields
  \begin{align} \label{EEPPSS}  \|(r^{(p)})_{ S \backslash \widehat{S}^{(p+1)}}\|_2^2
    =   \|(r^{(p)})_{ S \backslash \widetilde{L}}\|_2^2
      + \|(r^{(p)})_{ S \cap \widetilde{L}}\|_2^2.  \end{align}
By (\ref{ZZ01}), (\ref{OP-01}) and (\ref{EEPPSS}), one has
  \begin{align} \label{Bound11}  \Omega^2 & :=   \|(r^{(p)})_{(S\cup S^{(p)})\backslash \widehat{S}^{(p+1)} }\|_2^2
     =  \|(r^{(p)})_{ S \backslash \widehat{S}^{(p+1)}}\|_2^2\nonumber\\
   &  =     \|(r^{(p)})_{ S \backslash \widetilde{L} }\|_2^2 + \|(r^{(p)})_{ S  \cap  \widetilde{L} }\|_2 ^2   \nonumber\\
&  \leq  \|(r^{(p)})_{\widetilde{L}  \backslash  S }\|_2^2 +   \|(r^{(p)})_{ S  \cap  \widetilde{L} }\|_2^2  \nonumber \\
& = \|(r^{(p)})_{\widetilde{L}  \backslash (S\cup S^{(p)})}\|_2^2
  +   \|(r^{(p)})_{ S  \cap  \widetilde{L} }\|_2^2,
  \end{align}
  where the last equality follows from (\ref{OP-01}). In fact, by eliminating the components of $(r^{(p)})_{\widetilde{L}  \backslash  S } $ indexed by $S^{(p)},$ we immediately see that
   \begin{equation} \label{EQU22} \|(r^{(p)})_{\widetilde{L}  \backslash  S }\|_2 = \|(r^{(p)})_{\widetilde{L}  \backslash (S\cup S^{(p)})}\|_2. \end{equation}

   We now estimate the last term $ \|(r^{(p)})_{ S  \cap  \widetilde{L} }\|_2^2 $ in (\ref {Bound11}).
  Note that the elements in $ \Theta^{(p)}$ are added to $ S^{(p)}$ to form the next   set $ S^{(p+1)}. $  From the definition of $ \Theta^{(p)},$  it contains the indices in $ \widetilde{L} = L_k (r^{(p)})$ such that
 $ |(r^{(p)})_i| \geq \gamma \|r^{(p)}\|_\infty.$
 Clearly,  we have  either $ |\Theta^{(p)}| < k$ or $ |\Theta^{(p)}| = k. $  We  now consider the two cases separately.

 \textbf{\emph{Case 1:}}  $ |\Theta^{(p)}| = k. $ In this case, all indices of $ \widetilde{L}$ are selected into $ \Theta^k.$ So $ \Theta^{(p)} = \widetilde{L}.$ Under the assumption (\ref{COND0001}), it implies that $ S \cap  \widetilde{L} = S\cap \Theta^{(p)}=\emptyset$ and hence $\|(r^{(p)})_{S \cap \widetilde{L} } \|_2 =0.$ As a result, the relation (\ref{Bound11}) reduces to
 $ \Omega \leq   \|(r^{(p)})_{\widetilde{L}\backslash  (S \cup S^{(p)})} \|_2. $

 \textbf{\emph{Case 2:}} $ |\Theta^{(p)}| < k. $ In this case, by the structure of DOMP, not all indices in $ \widetilde{L} $ are selected into $\Theta^{(p)} .$  Without loss of generality, we assume that $ S\cap \widetilde{L} \not=\emptyset$ since otherwise $ \|(r^{(p)})_{ S  \cap  \widetilde{L} }\|_2^2 =0.$  Under the condition (\ref{COND0001}), we see that $ (S \cap \widetilde{L}  ) \cap \Theta^{(p)} = (S \cap \Theta^{(p)}) \cap \widetilde{L}=  \emptyset. $
 So the indices  in $ S \cap \widetilde{L} $ are not in $\Theta^{(p)}$, and hence
 $|(r^{(p)})_i| <  \gamma \|r^{(p)}\|_\infty ~ \textrm{for all } i \in S\cap \widetilde{L}, $ which implies that
  \begin{equation} \label{CCEE}  \|(r^{(p)})_{S \cap \widetilde{L}}\|_\infty < \gamma \|r^{(p)}\|_\infty.
  \end{equation}
 Again, as $ \Theta^{(p)} \subseteq \widetilde{L} $ and  the elements in $ S \cap \widetilde{L} $ are not in $\Theta^{(p)},$ we see that   $ \Theta ^{(p)} \subseteq \widetilde{L} \backslash ( \widetilde{L} \cap S) =\widetilde{L} \backslash S . $  Since the index of the largest magnitude  of   $r^{(p)}$ is in $\Theta^{(p)},$ one has
     $$\|r^{(p)} \|_\infty =  \|  (r^{(p)})_{\Theta^p}\|_\infty  \leq \|  (r^{(p)})_{\Theta^p}\|_2 \leq   \|(r^{(p)})_{\widetilde{L} \backslash  S}  \|_2. $$
    Merging this  relation and (\ref{EQU22})  yields
    \begin{equation} \label{CCEE1} \|r^{(p)} \|_\infty \leq \|(r^{(p)})_{\widetilde{L} \backslash  (S \cup S^{(p)})}  \|_2. \end{equation}
By using   (\ref{CCEE}) and (\ref{CCEE1}), we deduce that
  \begin{align} \label{EERR11}   \|(r^{(p)})_{S \cap \widetilde{L} } \|_2     &  \leq \sqrt{|S \cap \widetilde{L} |} \|(r^{(p)})_{S \cap  \widetilde{L} }  \|_\infty \nonumber\\
   & <  \gamma \sqrt{|S \cap \widetilde{L}|} \|r^{(p)}\|_\infty   \nonumber\\
   &  \leq \gamma \sqrt{|S \cap \widetilde{L}|} \|(r^{(p)})_{\widetilde{L} \backslash  (S \cup S^{(p)})} \|_2  \nonumber\\
      & \leq  \gamma \sqrt{k}  \|(r^{(p)})_{\widetilde{L}\backslash  (S \cup S^{(p)})} \|_2,
  \end{align}
 where the last inequality follows from  $  \gamma \sqrt{|S \cap \widetilde{L}|} \leq \gamma \sqrt{k}  .$
 Combining (\ref{Bound11}) and (\ref{EERR11}) leads to
 \begin{equation} \label{B-Case-1}  \Omega \leq \sqrt{1+ \gamma^2k}  \left\|(r^{(p)})_{\widetilde{L}\backslash  (S \cup S^{(p)})} \right\|_2. \end{equation}
So the bound (\ref{B-Case-1}) is valid in both Case 1 and Case 2.
Since  $ [\widetilde{L} \backslash (S\cup S^{(p)}) ] \cap \textrm{supp}(x_S-x^{(p)}) =\emptyset,  $ by Lemma \ref{Lem-Basic} (ii) and (iii), we obtain
\begin{align*}   &   \|(r^{(p)})_{\widetilde{L} \backslash (S\cup S^{(p)})} \|_2  \\
  & = \|(A^T A(x_S- x^{(p)})+ A^T\nu')_{\widetilde{L} \backslash (S\cup S^{(p)})} \|_2 \\
& \leq \|(A^T A(x_S- x^{(p)}))_{\widetilde{L} \backslash (S\cup S^{(p)})} \|_2 +  \|(A^T\nu')_{\widetilde{L} \backslash (S\cup S^{(p)})} \|_2 \\
& \leq \delta_{k+ |S^{(p)}|+|\widetilde{L}|} \|x_S- x^{(p)}\|_2+ \sqrt{1+\delta_k} \|\nu'\|_2\\
& = \delta_{2k+ |S^{(p)}| } \|x_S- x^{(p)}\|_2+ \sqrt{1+\delta_k}\|\nu'\|_2.
\end{align*}
Merging (\ref{B-Case-1}) and the relation above leads to
\begin{align}\label{Final-ER}  \Omega    \leq   \sqrt{1+ k\gamma^2 } \left(\delta_{2k+ |S^{(p)}| } \|x_S- x^{(p)}\|_2+  \sqrt{1+\delta_k} \|\nu'\|_2\right) .
\end{align}
Denote by
 $$ W :=  [(x_S-x^{(p)}) -  A^T (y-Ax^{(p)}) ]_{(S\cup S^{(p)}) \setminus  \widehat{S}^{(p+1)} } . $$
Note that $|\textrm{supp} ( x_S-x^{(p)})| \leq k+|S^{(p)}|  $ and $|(S\cup S^{(p)}) \setminus \widehat{S}^{(p+1)} | \leq |S| =k$ since $ S^{(p)} \subseteq \widehat{S}^{(p+1)}. $  By Lemma \ref{Lem-Basic} (i) and (iii), one has
\begin{align}\label{WWW111}   \|W\|_2
 & = \| [(I - A^TA) (x_S-x^{(p)}) - A^T \nu']_{(S\cup S^{(p)})\backslash \widehat{S}^{(p+1)} } \|_2  \nonumber \\
&   \leq \delta_{ k + |S^{(p)}|} \|x_S-x^{(p)}\|_2+\sqrt{1+\delta_k} \|\nu'\|_2.
 \end{align}
Note that $ x_S-x^{(p)}= (x_S    -x^{(p)})_{(S\cup S^{(p)})}$  since $ \textrm{supp} (x_S-x^{(p)}) \subseteq S\cup S^{(p)}. $ We have
 \begin{align} \label{OOO111}      \|(x_S    &    -x^{(p)})_{\overline{\widehat{S}^{(p+1)} }}\|_2 \nonumber \\
  & = \| (x_S    -x^{(p)})_{(S\cup S^{(p)}) \backslash \widehat{S}^{(p+1)}}\|_2  \nonumber \\
  & = \| [x_S-x^{(p)} -  A^T (y-Ax^{(p)})]_{(S\cup S^{(p)}) \backslash \widehat{S}^{(p+1)}}  \nonumber \\
  & ~~~ + [A^T (y-Ax^{(p)})]_{(S\cup S^{(p)}) \backslash \widehat{S}^{(p+1)}}  \|_2  \nonumber \\
 & \leq \|W\|_2+ \Omega.
\end{align}
  Denote by $ \phi: =1+\sqrt{1+ k\gamma^2}. $   Combining (\ref{Final-ER}), (\ref{WWW111}) and (\ref{OOO111}) yields
\begin{align*}       \|(x_S    &   -x^{(p)})_{ \overline { \widehat{S}^{(p+1)} }} \|_2 \\
& =  \left(\delta_{ k +|S^{(p)}|}    +   \sqrt{1+ k\gamma^2} \delta_{2k+| S^{(p)}|}  \right) \|x_S- x^{(p)}\|_2  \\
 & ~~ +     \phi  \sqrt{1+\delta_k} \|\nu'\|_2\\
& \leq       \phi \left( \delta_{2k+| S^{(p)}|}  \|x_S- x^{(p)}\|_2
 +       \sqrt{1+\delta_k} \|\nu'\|_2\right),
\end{align*}
where the last inequality follows from the fact $ \delta_{ k +|S^{(p)}|} \leq \delta_{2k+| S^{(p)}|}. $ Thus (\ref{EELL02}) is satisfied.
  \hfill $ \Box $

We have all ingredients needed to show the main result stated in Theorem \ref{Thm-DOMP}. \\

\textbf{\emph{The proof of Theorem \ref{Thm-DOMP}:}} Let $ x^{(p)}$ be  the current iterate and  $S^{(p)}, \Theta^{(p)}, S^{(p+1)} $ and  $ x^{(p+1)}$ are defined by DOMP. Without loss of generality, we assume $ r^{(p)}\not=0. $    Let $ \widehat{S}^{(p+1)}$ be defined as (AC1) and $ \widehat{x}^{(p+1)} $ be a solution to the auxiliary  minimization problem in (AC2). Still,  $ S =L_k(x)$ is the index set for the $k$ largest magnitudes of the target vector $ x.$
At the $p$th iteration, there are only two cases: either $ S\cap \Theta^{(p)}  \not=\emptyset$ or $ S\cap \Theta^{(p)}   =\emptyset. $

\emph{Case 1:} $S\cap \Theta^{(p)}  \not=\emptyset.$ In this case, at least one of the indices in $S$ is added to $S^{(p+1)} = S^{(p)} \cup \Theta^p .$   By (\ref{CSSS}),  $ \Theta^{(p)} \cap S^{(p)} = \emptyset. $ Thus set  $S^{(p+1)}  $ contains more elements of $S$ than $ S^{(p)}$ in this case.  Note that $\textrm{supp} (x^{(p)}) \subseteq S^{(p)}  \subseteq  S^{(p+1)}. $  Thus $x^{(p)}$ is a feasible vector to the optimization problem (DOMP2), to which $ x^{(p+1)}$ is a minimizer. Thus by optimality, one has
$ \|y-Ax^{(p+1)}\|_2 \leq \|y-Ax^{(p)}\|_2. $
Note that $ y= Ax_S + \nu' $ where $ \nu' = A x_{\overline{S}} +\nu. $ By the triangular inequality,  it follows from the inequality above that
 $$ \|A(x_S-x^{(p+1)})\|_2 - \|\nu'\|_2  \leq \|A(x_S-x^{(p)})\|_2 +\|\nu'\|_2, $$
 By noting that $ |\textrm{supp} (x_S-x^{(p+1)})| \leq k+|S^{(p+1)}| $ and $|\textrm{supp}( x_S-x^{(p)})| \leq k+|S^{(p)}|  $ and by the definition of RIC of $A$,
 the inequality above implies that
 \begin{align*}      \sqrt{1-\delta_{k+|S^{(p+1)}|}}  &    \|x^{(p+1)}- x_S\|_2 \\
  &  \leq   \sqrt{ 1+\delta_{k+|S^{(p)}|} }   \|x^{(p)}- x_S\|_2   +  2\|\nu'\|_2. \end{align*}
Since  $\delta_{k+|S^{(p)}|} \leq  \delta_{k+|S^{(p+1)}|} ,$  one has
\begin{align} \label{CASE1} \|x^{(p+1)}- x_S\|_2  \leq  &  \sqrt{ \frac{1+\delta_{k+|S^{(p+1)}|} } { 1-\delta_{k+|S^{(p+1)}|}  }  }  \|x^{(p)}- x_S\|_2 \nonumber \\
 &    +  \frac{2}{\sqrt{1-\delta_{k+|S^{(p+1)}|}} }\|\nu'\|_2.
\end{align}

 The first relation in (\ref{DDTT}) implies that  $ |S^{(p+1)}| = \sum_{i=1}^p |\Theta^{(i)}|  , $ which means $\{ |S^{(p)}|\}$ is nondecreasing sequence.
Let $p^*$  denote the largest integer number $ p$  such that $ |S^{(p+1)}|   \leq  (c-2)k.$   Then for any $ p\leq p^*,$ the coefficients  on the right-hand side of (\ref{CASE1}) are bounded by
 \begin{equation} \label{rrr}   \sqrt{ \frac{1+\delta_{k+|S^{(p+1)}|} } { 1-\delta_{k+|S^{(p+1)}|}  }  }  \leq \varrho,
~~  \frac{2}{\sqrt{1-\delta_{k+|S^{(p+1)}|}} } \leq c_1, \end{equation}    where $ \varrho:  = \sqrt{ \frac{1+\delta_{ck} } { 1-\delta_{ck}  }  }$ and   $ c_1:= \frac{2}{\sqrt{1-\delta_{ck}}}. $   Thus it follows from
(\ref{CASE1}) and (\ref{rrr}) that
\begin{equation}
\label{CASE1-err}  \|x^{(p+1)}- x_S\|_2  \leq  \varrho  \|x^{(p)}- x_S\|_2 +   c_1\|\nu'\|_2
 \end{equation}
  for any $p \leq   p^*. $  We now analyze the second case.

\emph{Case 2:}  $ S\cap \Theta^{(p)} =\emptyset.$ Let $ p^*$ be defined as above, i.e., $ p^*$ is the largest number $p$ such that $|S^{(p+1)}| \leq (c-2)k.$  Denote by $ \phi =  1 + \sqrt{1+k\gamma^2}. $
By Lemmas \ref{Lem-ERR-01} and  \ref{LEM-Final}, we immediately obtain that
{\small \begin{align} \label{IN0011}  & \|x_S     -\widehat{x}^{(p+1)} \|_2   \nonumber  \\
   &  \leq \frac{   \phi(  \delta_{2k+ |S^{(p)}|}    \| x_S-x^{(p)}\|_2
      +   \sqrt{1+\delta_k} \|  \nu'\|_2   )  }  {\sqrt{1- \delta^2_{k+|\widehat{S}^{(p+1)}| } } }  \nonumber  \\
       & ~~    + \frac{ \sqrt{1+\delta_{| \widehat{S}^{(p+1)} | }}  \| \nu'   \|_2}{ 1-\delta_{k+|\widehat{S}^{(p+1)}| } } \nonumber  \\
      & =  \frac{  \phi \delta_{2k+ |S^{(p)}|} \| x_S-x^{(p)}\|_2  }  {\sqrt{1- \delta^2_{k+|\widehat{S}^{(p+1)}| } } }  \nonumber  \\
               & ~~ +    \left(\frac{ \phi  \sqrt{1+\delta_k}  }  {\sqrt{1- \delta^2_{k+|\widehat{S}^{(p+1)}| } } }   + \frac{\sqrt{1+\delta_{| \widehat{S}^{(p+1)} | }}} { 1-\delta_{k+|\widehat{S}^{(p+1)}| } } \right)  \|  \nu'   \|_2 \nonumber \\
      & \leq  \frac{ \phi \delta_{2k+ |S^{(p+1)}|} \| x_S-x^{(p)}\|_2  }  {\sqrt{1- \delta^2_{2k+ |S^{(p+1)}|}   } }  \nonumber  \\
        & ~~   +    \left(\frac{ \phi \sqrt{1+\delta_k} }  {\sqrt{1- \delta^2_{2k+ |S^{(p+1)}|}   } }   + \frac{\sqrt{1+\delta_{ 2k+ |S^{(p+1)}|  }}} { 1-\delta _{2k+ |S^{(p+1)}|}  } \right)  \| \nu'   \|_2  \nonumber\\
          & \leq  \frac{ \phi \delta_{ck} \| x_S-x^{(p)}\|_2  }  {\sqrt{1- \delta^2_{ck}   } }
          +    \left(\frac{ \phi \sqrt{1+\delta_k }  }  {\sqrt{1- \delta^2_{ck}  } }   + \frac{ \sqrt{1+\delta_{ck} }  }{ 1-\delta _{ck}  } \right)  \|  \nu'   \|_2,
     \end{align} }
 where the second inequality follows from the fact that $ \delta_{2k+|S^{(p)}|} \leq \delta_{2k+|S^{(p+1)}|} $ since $
     |S^{(p)}| \leq |S^{(p+1)}| $, and that $\delta_{k+|\widehat{S}^{(p+1)}|} \leq \delta_{2k+|S^{(p+1)}|} $ since $
     |\widehat{S}^{(p+1)}| = |S^{(p+1)}| +|\Lambda^{(p)}|  \leq k+ |S^{(p+1)}|$  which follows from the second relation in (\ref{DDTT}).   The last inequality follows from the fact $ \delta_{2k+|S^{(p+1)}|} \leq \delta_{ck} $ due to  $ p\leq p^*.$
Denote by
\begin{equation} \label{EEERRR3} \epsilon:= \frac{\vartheta_{(y,A)} }{\sqrt{1-\delta_{ck}} }\left( \sqrt{\left(\frac{2} {\sqrt{\sigma}}  + \frac{\|A\|_2} { \sigma }  \right)  \| A\|_2     }  + \frac{\|A\|_2}{\sqrt{\sigma}} \right) ,
\end{equation}
 where $ \vartheta_{(y,A)} $ and $\sigma $ are defined in Lemma \ref{proximity}.
  Since $ \textrm{supp}( \widehat{x}^{(p+1)}-x^{(p+1)}) \subseteq \widehat{S}^{(p+1)}, $ by the definition of RIC of $A$ and Lemma  \ref{proximity}, we have
\begin{align} \label{INaabb} &  \|\widehat{x}^{(p+1)}     -x^{(p+1)}\|_2  \nonumber \\
&\leq \frac{\| A(\widehat{x}^{(p+1)}-x^{(p+1)})\|_2}{\sqrt{1-\delta_{|\widehat{S}^{(p+1)}|}}} \leq \frac{\| A(\widehat{x}^{(p+1)}-x^{(p+1)})\|_2 }{\sqrt{1-\delta_{ck}}} \nonumber \\
&\leq   \frac{\vartheta_{(y,A)} }{\sqrt{1-\delta_{ck}} }\left( \sqrt{\left(\frac{2} {\sqrt{\sigma}}  + \frac{\|A\|_2} { \sigma }  \right)  \| A\|_2     }  + \frac{\|A\|_2}{\sqrt{\sigma}} \right) \nonumber \\
&   = \epsilon .
\end{align}
  Then
using the triangular inequality and (\ref{IN0011}) and (\ref{INaabb}) yields
  \begin{align}  \label{CASE2-err}
    \|x _S       -x^{(p+1)}\|_2
  & \leq \|x_S -\widehat{x}^{(p+1)}\|_2 + \|\widehat{x}^{(p+1)} - x^{(p+1)}\|_2 \nonumber \\
    & \leq    \frac{  \phi \delta_{ck} }  {\sqrt{1- \delta^2_{ck}} }
          \| x_S-x^{(p)}\|_2   \nonumber \\
          &   ~~   +  \left( \frac{\phi \sqrt{1+\delta_k} }  {\sqrt{1- \delta^2_{ck} } }   + \frac{\sqrt{1+\delta_{ck} }}{ 1-\delta_{ck} } \right) \|\nu'\|_2+ \epsilon    \nonumber \\
      & =  \beta  \| x_S-x^{(p)}\|_2  + c_2  \| \nu'\|_2+ \epsilon,
\end{align}
where the constants $\beta $ and $c_2 $ are given by
\begin{equation} \label{5566}      \beta = \frac{  \phi \delta_{ck} }  {\sqrt{1- \delta^2_{ck}} } ,
 ~~  c_2 =    \left(\frac{ \phi \sqrt{1+\delta_k} }  {\sqrt{1- \delta^2_{ck} } }   + \frac{\sqrt{1+\delta_{ck}}}{ 1-\delta_{ck} } \right) .     \end{equation}
 It is easy to verify that $ \beta <1$ under the assumption of the theorem.  In fact,
 $  \beta= \frac{    \phi  \delta_{ck}  }  {\sqrt{1- \delta^2_{ck} } } <1  $  is guaranteed under the condition
   $$ \delta_{ck} <  \frac{1}{ \sqrt{1+\phi^2}} =   \frac{1} {\sqrt{ 1+ \left(1+ \sqrt{ 1+ k\gamma^2}\right )^2  }} . $$

Suppose that the algorithm DOMP has performed $p$ iterations where $ p\leq p^*. $ Without loss of generality, suppose that within these $p$ iterations the aforementioned \emph{Case 1} appears $\tau^*$ times, and thus \emph{Case 2} appears $ p-\tau^*$ times. Clearly,  \emph{Case 1} appears at most $k$ times since $ |S| =k,$  so $ \tau^* \leq k. $ The error bound for this case is given by (\ref{CASE1-err}),  and   the error bound for \emph{Case 2} is given by (\ref{CASE2-err}).
 Therefore, it is not difficult to verify that after $p$ iterations, one has
 \begin{align} \label{Combined-IN}     \|x_S-x^{(p+1)}\|_2     \leq  &    (\varrho^{\tau^*} \beta^{p-\tau^*})\|x_S-x^{(0)}\|_2 \nonumber \\
  & + c_1\beta^{p-\tau^*} \left( \frac{\varrho^{\tau^*} -1}{\varrho-1}\right)\|\nu'\|_2 \nonumber \\
     & +  \frac{ c_2\|\nu'\|_2  + \epsilon}  {1-\beta}     .
 \end{align}
 where all constants $(\beta, \varrho, c_1, c_2) $ are  determined only by $  \delta_{ck} $ and/or $ k$ and $ \gamma . $
 To see (\ref{Combined-IN}), without loss of generality, we assume that the first $\tau^*$ iterations correspond  to  \emph{Case 1}, and the remaining $ p-\tau^* $ iterations correspond to \emph{Case 2}.
 Then using the fact $\sum_{i=0}^ {p-\tau^*-1} \beta^i = (1- \beta^{p-\tau^*})/(1-\beta) < 1/(1-\beta) $ for  $ \beta <1,$  it follows from (\ref{CASE2-err})  that
 \begin{align} \label{FFFI01}   \|x_S     -x^{(p+1)}\|_2
   & \leq \beta^{p-\tau^*} \|x_S-x^{(\tau^*+1)}\|_2 \nonumber \\
    & ~~ + (1+ \beta + \dots + \beta^{p-\tau^*-1}) (c_2 \|\nu'\|_2 + \epsilon) \nonumber  \\
 & \leq \beta^{p-\tau^*} \|x_S-x^{(\tau^*+1)}\|_2 + \frac{c_2 \|\nu'\|_2 + \epsilon}{ 1-\beta}. \end{align}
Also, by  (\ref{CASE1-err}) and the fact $\sum_{i=0}^{\tau^*-1}  \varrho^i  =  (\varrho ^{\tau^*}-1)/(\varrho-1) $ where $ \varrho>1,$ we obtain
   \begin{align}  \label{FFFI02}       \|x_S   &  -x^{(\tau^* +1)}\|_2   \nonumber \\   & \leq \varrho^{\tau^*} \|x_S-x^{(0)}\|_2 +  (1+ \varrho + \dots +  \varrho^{\tau^*-1}) (   c_1  \|\nu'\|_2 ) \nonumber  \\
     &  =  \varrho^{\tau^*} \|x_S-x^{(0)}\|_2 + c_1 \left( \frac{\varrho^{\tau^*} -1}{\varrho-1}\right)\|\nu'\|_2.
   \end{align}
  Merging  (\ref{FFFI01}) and (\ref{FFFI02})  yields
 \begin{align*}    \|x_S    &    -x^{(p+1)}\|_2      \leq        \beta^{p-\tau^*}(\varrho^{\tau^*}\|x_S-x^{(0)}\|_2)  \nonumber \\
       &    + c_1\beta^{p-\tau^*} \left( \frac{\varrho^{\tau^*} -1}{\varrho-1}\right)\|\nu'\|_2 +  \frac{ c_2\|\nu'\|_2  + \epsilon}  {1-\beta} ,
\end{align*}
which is exactly the error bound given in (\ref{Combined-IN}).
Note that the coefficients $  (\varrho/\beta)^{\tau^*} \leq (\varrho/\beta)^{k} $  and  $ c_1\beta^{p-\tau^*} \left( \frac{\varrho^{\tau^*} -1}{\varrho-1}\right) \leq  c_1  \left( \frac{\varrho^k -1}{\varrho-1}\right)$ since $ \beta   <1,$ $\varrho >1 $  and $  \tau^*  \leq k .$
 Thus we deduce that
  \begin{align} \label{WWW00}    \|x_S     -x^{(p+1)}\|_2     &   \leq    \beta ^p \left[(\varrho/\beta)^{k} \|x_S-x^{(0)}\|_2\right]
  +  \tau \|\nu'\|_2  \nonumber \\
     & ~~ + \epsilon /(1-\beta). \end{align}
  where
 $ \tau   =c_1  \left( \frac{\varrho^{k} -1}{\varrho-1}\right) + \frac{ c_2}{1-\beta}  $  with constants $ c_1 =2/\sqrt{1-\delta_{ck}}$ and $c_2$ given in (\ref{5566}).
The term $(\varrho/\beta)^k \|x_S-x^{(0)}\|_2 $ in (\ref{WWW00}) is a fixed quantity. In (\ref{WWW00}), only the last term $ \epsilon/(1-\beta)$ depends on $ \sigma $, and we see from the definition of $ \epsilon$ that $ \epsilon \to 0$ as $ \sigma \to \infty. $ Therefore, the desired estimation (\ref{FERR}) is obtained.
    \hfill $\Box $

The  main  result  established in this section provides an estimation for the distance between the significant components of the target data and   the solution generated by the DOMP algorithm. The result claims that the reconstruction quality becomes better and better as the iteration proceeds provided that   $ |S^{(p)}| \leq  (c-2)k,$ where $ k$ is the sparsity level of the target data and  $ ck $   is the  order of the RIC of the measurement matrix.  In this process, the algorithm gradually identifies the correct support of the sparse data and the iterate $x^{(p)}$ gets closer and closer to the significant components of the target data. When $ m,n $ are particularly large with $ n \gg m$ and the highest order RIC of $ A$ is much larger than $ 2k, $ the proposed algorithms can perform enough iterations to guarantee that the reconstruction error  becomes so small that the finite convergence of the algorithms can be achieved,  as discussed in the next subsection.

\subsection{Finite convergence}\label{sect05}

From Definition \ref{Def1.1},  we see  that  $ \delta_s \leq \delta_t $ for any positive integer number  $ s\leq t.$ Thus if $A$ has the RIP of order $t$, i.e.,  $ \delta_t <1, $  then it has the RIP of any order $ s \leq t. $ In this section, we define the  highest order of the RIP of $A$ as $$ t^{\max} = \max \{t:  ~\delta_{t} <1\}. $$
The basic assumption in compressed sensing is that the number of measurements  $m$  is much lower than the signal length $n,$ i.e.,  $ m \ll n.  $  It is also well known that to ensure a  $k$-sparse signal being exactly reconstructed, the signal should be sparse enough so that it is the unique sparsest solution to the underdetermined linear system $y= Ax  $  (see \cite{E10}). We now consider the reconstruction problems satisfying the following assumption.

\begin{Assump} \label{Aum1}
$ k \ll m \ll n $ and $ 2k \ll t^{\max}, $ where $t^{\max}$ is  the highest order of RIP of the measurement matrix $A.$
\end{Assump}

 The problem in such settings  may be referred to as a large-scale compressed sensing problem for which the DOMP algorithm can maintain the error bound (\ref{FERR}) with enough iterations so that the significant information of the target data can be guaranteed to reconstruct. Let $ c$ be the largest integer number such that $ 2k  \ll ck\leq t^{\max},$ i.e., $ c=\lfloor \frac{t^{\max}}{k}  \rfloor. $ Recall that $ p^*$ (defined in the proof of Theorem \ref{Thm-DOMP}) denotes  the largest number of iterations such that $ |S^{(p^*+1)}| \leq (c-2)k,$ i.e.,
  $  \sum_{i=1}^{p^*} | \Theta^{(i)} | \leq (c-2)k. $  This means that $  \sum_{i=1}^{p^*+1} | \Theta^{(i)} | >  (c-2)k $ which together with  $ | \Theta^{(i)} | \leq k$  implies that $ (p^*+1) k > (c-2)k$, i.e.,  $ p^* \geq c-2.$ Thus $ p^*$ would be a large number when $ c \gg 2  $ which is guaranteed by $ t^{\max}\gg 2k . $

  We now point out that for  large-scale compressed sensing problems, the DOMP algorithm can exactly reconstruct the support of $ x_S $  in a finite number of iterations as long as $\|Ax_{\overline{S}}+\nu\|_2 $ is small enough.  We say that the vector $x$ is \emph{nondegenerated $k$-compressible} if it has at least $k$ nonzero components and $\|x_{\overline{S}}\|_2$ is sufficiently small, where $ S= L_k(x).$ For such a vector, $ x_i \not= 0 $  for all $ i\in S  .$ Clearly, any $k$-sparse vector with $k$ nonzeros must be nondegenerated $k$-compressible.

 \begin{Thm} \label{asymptotic}  Assume that  the data $ x\in \mathbb{R}^n $ is nondegenerated $k$-compressible.  Let $y:=Ax+\nu $ be the slightly inaccurate measurements of  $x $ in  the sense that the errors $ \nu $  are sufficiently small. Let $ \gamma \in(0,1] $ be a given constant in DOMP. Suppose that Assumption \ref{Aum1} holds and that $A$ has the RIP of order $ ck  $  with $t^{\max} \geq ck \gg  2k$ and $\delta_{ck}$  satisfies (\ref{FERR}). Then when $ p^*= \max\{p: |S^{(p+1)}|\leq (c-2) k\}$ is large enough, there must exist $ \widehat{p} $ such that for any $ p$ with $ \widehat{p}  \leq p \leq p^*, $  the iterate $ x^{(p)}$ generated by DOMP satisfies that
 $$ L_k( x^{(p)}) =L_k(x)  . $$

\end{Thm}

 \emph{Proof.} (i)  Under (\ref{FERR}),  Theorem \ref{Thm-DOMP} claims that\begin{equation} \label{EERRFF} \|x^{(p)}- x_S \|_2 \leq \beta ^{p -1} (\frac{\varrho}{\beta})^k \|x^0-x_S\|_2 + \tau \| \nu'\| ,  \end{equation}
   where $S=L_k(x)$  and  the constants $(\beta, \varrho, \tau) $ are given in Theorem \ref{Thm-DOMP}.
 By assumption, the right-hand side of the above  bound is sufficiently small when $ p$ is large enough. Note that  $\|x^{(p)}- x_S \|_2^2 =\|(x^{(p)})_S- x_S \|_2^2+ \|(x^{(p)} )_{\overline{S}} \|_2^2.$ It follows from (\ref{EERRFF}) that both terms $ \|(x^{(p)})_S- x_S \|_2$ and $ \|(x^{(p)} )_{\overline{S}} \|_2 $ are sufficiently small when $ p$ is large enough, which is ensured under the assumption of the theorem.    Since $x$ is `nondegenerated $k$-compressible', this implies that the largest $k$ magnitudes of $ x^{(p)} $ are concentrated on $S$ and sufficiently close to $ x_S.$   This means $ L_k(x^{(p)}) =S. $ Thus after performing enough iterations, the support of $ x_S$ is correctly reconstructed by DOMP.
 \hfill $ \Box $

The error bound established in this paper measures  how close the iterates generated by the algorithm to the target data in terms of $ \delta_{ck},$   the number of iterations $p,$  and the measurement error $ \|\nu\|_2.$ The bound (\ref{FERR}) indicates  that as $ p$  increases, the reconstruction error decays provided that $ p \leq p^*.$ In the setting of large-scale compressed sensing, Theorem \ref{asymptotic} further claims that the exact reconstruction of the sparse data can be also possible under the assumption of the theorem.

\begin{Rem}  The EDOMP algorithm shares the same steps of DOMP when $ |S^{(p)}| \leq k, $ and performs hard thresholding to restore the $k$-sparsity of iterates when $ |S^{(p)}| > k. $ The iterate $ x^{(p+1)},$ generated by EDOMP when $ |S^{(p)}| > k, $  is very different from the one generated by DOMP. Clearly, the relations $ S^{(p)} \subseteq S^{(p+1)}$  and $|S^{(p)}|=\sum_{i=1}^p |\Theta^{(i)}|$ in DOMP are lost in EDOMP. Thus the analysis of DOMP presented in this section would not be generalized straightforwardly to EDOMP.  While we had attempted to show the reconstruction error bound for EDOMP in the earlier draft of this report, the argument therein for EDOMP remains incomplete. It is still not completely clear at the moment about the error bound for EDOMP. This is an interesting and worthwhile future work.
\end{Rem}

 \section{Numerical experiments}\label{sect5}

The performance of the proposed algorithms is clearly related to three main factors: (a) the choice of parameter $\gamma,$  (b) the relative sparsity level $k/m$  or $k/n$, and (c) the total number of iterations executed. We first  carry out experiments to demonstrate how these factors might affect the performance of the  proposed algorithms and thus to suggest the suitable choice of   $ \gamma$ as well as the number of iterations to perform.
 All sparse vectors and measurement matrices in experiments are randomly generated.
The entries of sparse vectors and matrices are assumed to be independent and identically distributed and follow the standard normal distribution, and the positions of nonzero entries of sparse vectors are uniformly distributed. All iterative  algorithms in our experiments use $x^{(0)}=0 $ as the initial point. Unless otherwise stated, we  adopt  the following reconstruction criterion:   \begin{equation} \label{CRC} \|x^{(p)}- x\|_2/\|x\|_2 \leq 10^{-5}, \end{equation}
where $ x^{(p)}$ is the iterate produced by the algorithm and $ x$ is the target data to reconstruct. If $ x^{(p)}$ satisfies (\ref{CRC}), we say that the algorithm succeeds in reconstructing $ x.$

 \subsection{Effect of the parameter $\gamma$}
 To illustrate how the value of $\gamma$ affects the reconstruction ability of the proposed algorithms, we compare the success rates of the proposed algorithms in $k$-sparse data reconstruction using the following 20 different values:  $ \gamma = t/20,$ where $ t=1,2, \dots,20.$    The size of matrix is $500 \times 2000,$ and the sparsity levels $k= 120, 140, 150, 160, 170 $ and $ 180$ of the sparse vectors $ x\in \mathbb{R}^{2000}$  are examined in this experiment.   For every given sparsity level $k,$  the success rates for $k$-sparse data reconstruction via  DOMP and EDOMP with each given  $\gamma$ are obtained respectively by attempting 500 random pairs $(A, x$) with accurate measurements $y:= Ax.$  The algorithms are performed  a total of $k$ iterations, where $ k$ is the sparsity level of the vector to reconstruct.
\begin{figure} [htp]
 $ \begin{array}{c}
\includegraphics [width=0.45\textwidth,
totalheight=0.225\textheight] {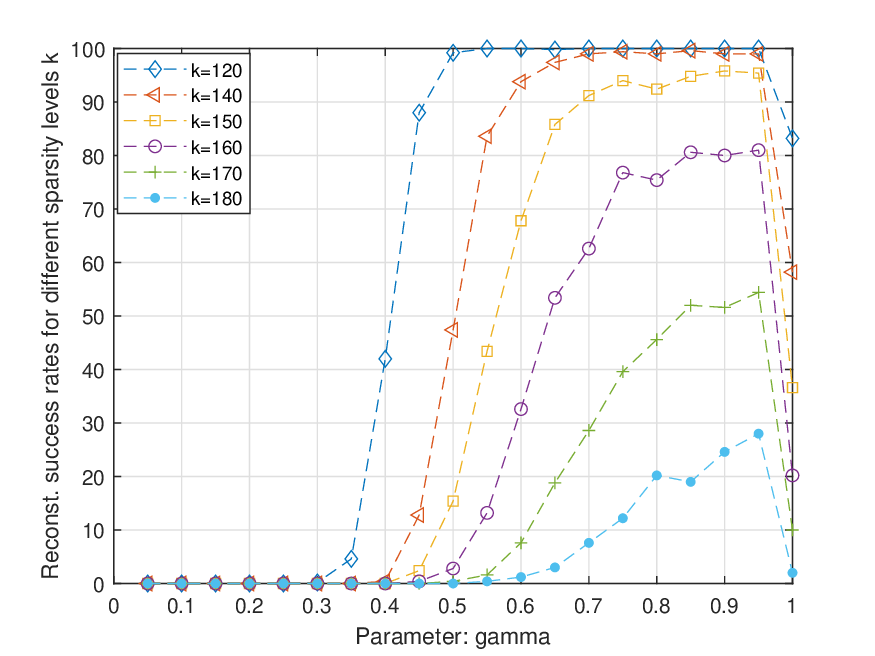} \\
  \textrm{(a) DOMP: Performance vs  $ \gamma$}
\end{array} $
 \hfill   $\begin{array}{c}
\includegraphics [width=0.45\textwidth,
totalheight=0.225\textheight] {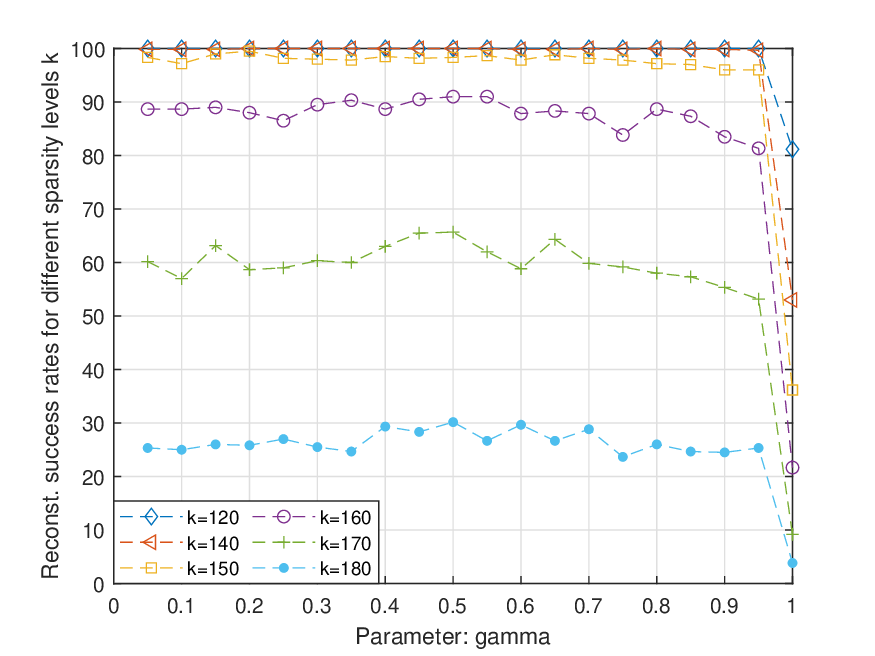} \\
\textrm{(b)  EDOMP: Performance vs  $ \gamma$}
\end{array}
$
 \caption{Comparison of performances of DOMP and EDOMP with different choices of $\gamma.$ } \label{figure1}
\end{figure}
 The results are given in Fig. \ref{figure1}.  Intuitively, a small value of $\gamma$ allows  more indices to be added to $\Theta^{(p)}$ at every iteration, but it may also increase the chance for  an incorrect index being added to $ \Theta^p$, and thus a small $\gamma $ may lower the success rates of DOMP for data reconstruction as shown in Fig. \ref{figure1}(a). However, as shown in Fig. \ref{figure1} (b), it seems that the performance of EDOMP is less sensitive to the change of $\gamma,$ compared with DOMP. On the other side,   the results in Fig. \ref{figure1} indicate that the success rates of the proposed algorithms with $ \gamma =1$ are not the highest ones. This does indicate that the algorithm with  $ \gamma=1$ (including OMP)  are not working to their potentials to identify the support of the target data. One can also see from Fig. 1(a) that the sparser the target data, the wider  range of values for $\gamma$ can be used in DOMP.   The numerical experiments indicate that the values of $\gamma$ situated between 0.7 and 1 are generally good for DOMP, while EDOMP admits more freedom in the choice of $ \gamma.$  As a result, we may use $ \gamma=0.9$ as a default choice for two algorithms in the remaining experiments.

 \subsection{Effect of the number of iterations} The main feature of OMP is that the support of the target data is gradually identified as the iteration proceeds.  This means that the support of the target data (which is usually unknown beforehand) cannot be fully obtained if the number of iterations performed is too low. Similarly, the proposed algorithms need to perform enough iterations before being terminated.  However, after the necessary iterations have been performed, can any further iterations continue to remarkably improve the performance of the proposed algorithms? To find a possible answer, it is useful to carry out an experiment to test the performance of algorithms against the number of iterations executed.
  \begin{figure} [htp]
    $\begin{array}{c}
\includegraphics [width=0.45\textwidth,
totalheight=0.225\textheight] {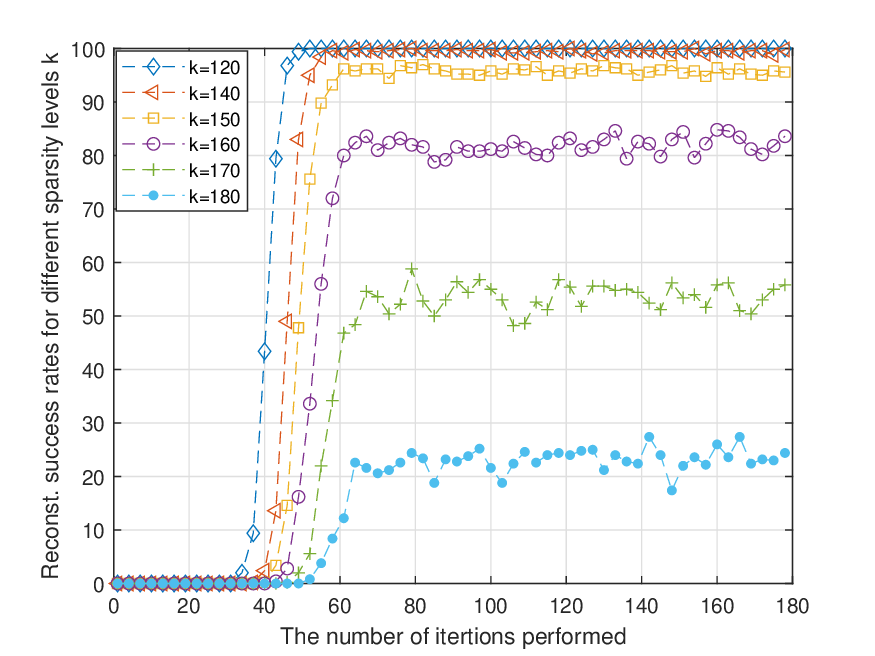} \\
\textrm{(a)  DOMP: Performance vs no. of iter. }
\end{array}
$
\hfill  $\begin{array}{c}
\includegraphics [width=0.45\textwidth,
totalheight=0.225\textheight] {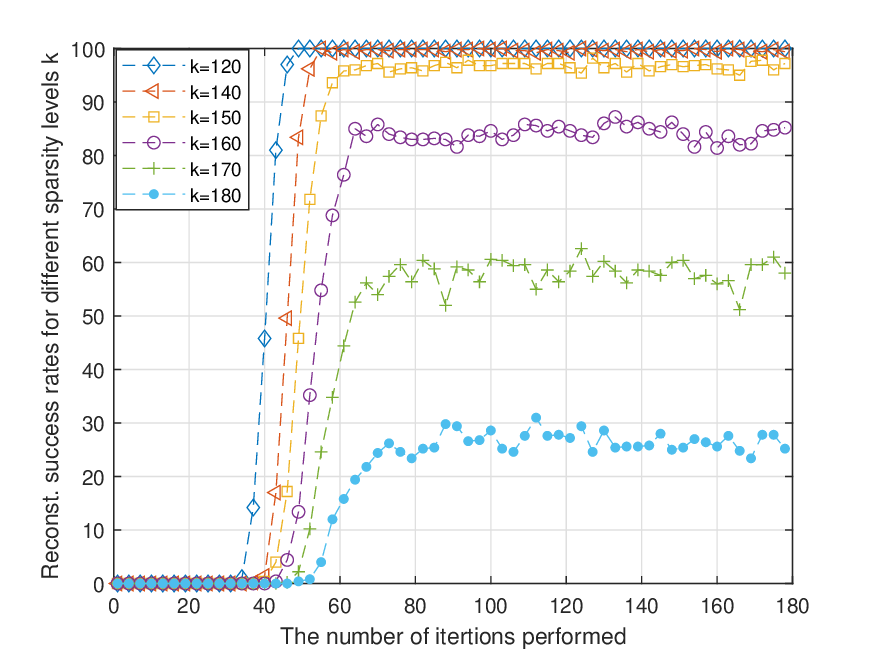} \\
\textrm{(b)  EDOMP: Performance vs no. of iter. }
\end{array}
$
 \caption{ Comparison of performances of DOMP and EDOMP with different no. of iterations.  } \label{figure2}
\end{figure}
According to our theory, under the RIP assumption, the DOMP and EDOMP algorithms keep  finding the correct support of the target data provided that the performed iterations are lower than $p^*$, the largest integer number $ p$ such that $|S^{(p+1)}| \leq (c-2)k$ where $ c>2$ is the constant associated with the order of RIP.  But when iterations go beyond $ p^*,$ there would be no guarantee for the algorithms to continue to find the correct support of $ x_S.$
With the same random matrices and sparse vectors used in first experiment, we compare the performance of DOMP and EDOMP when they are allowed to perform different numbers of iterations: $p= 1+3j$ where $j =0, 1, \dots, 59.$    For every given sparsity level $k=120, 140, 150, 160, 170, 180$ and the prescribed number for iterations,  the success rates for data reconstruction via the  two algorithms with $\gamma=0.9$  are obtained by 500 random tries of $(A,x)$ with accurate measurements.  The results are given in Fig. \ref{figure2}.  As expected, the algorithms need to perform  necessary iterations in order to possibly reconstruct the data. The figure indicates that with the increment of iterations, the success frequencies of the two algorithms for data reconstruction increase  to some highest point,  after that the  performance of algorithms would generally not be improved  no matter how many further iterations are performed. From Fig. \ref{figure2}, it can be easily seen that the highest point of the reconstruction curve appears after a certain number of iterations that is much lower than the sparsity level $k$ of the vector. This means unlike the classic OMP, the DOMP and EDOMP algorithms do not need to perform at least $k$ iterations in order to reconstruct a $k$-sparse signal.  Thus we carry out further simulations to examine the average number of iterations as well as the average runtime needed for the proposed algorithms to reconstruct sparse data.

 \subsection{Average number of iterations and average runtime}

   Experiments were performed to demonstrate the average number of iterations and runtime needed for the proposed algorithms to achieve the reconstruction  criterion (\ref{CRC}).
 \begin{figure} [htp]
 $ \begin{array}{c}
\includegraphics [width=0.45\textwidth,
totalheight=0.225\textheight] {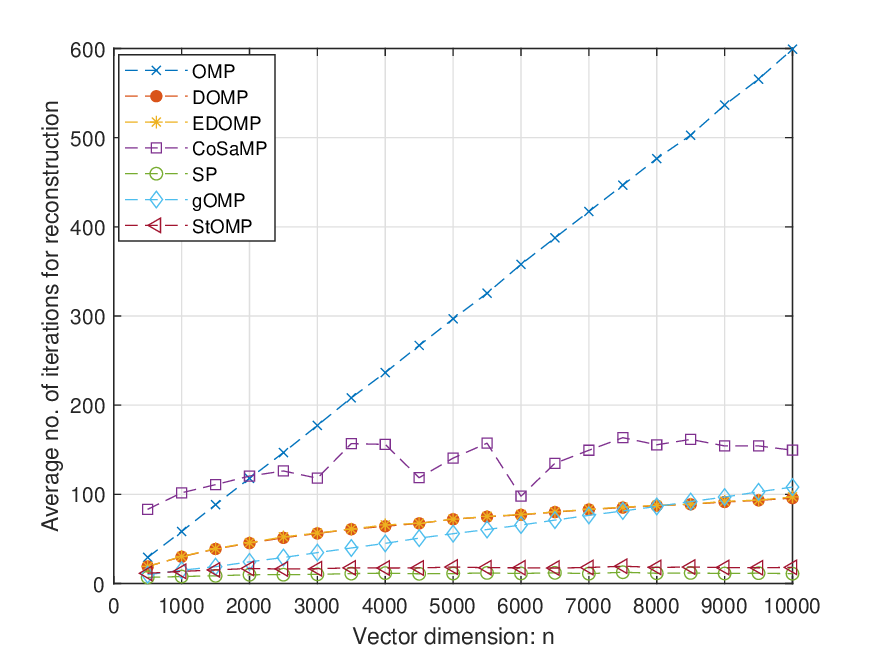} \\
  \textrm{(a) Average no. of iterations }
\end{array} $
 \hfill   $\begin{array}{c}
\includegraphics [width=0.45\textwidth,
totalheight=0.225\textheight] {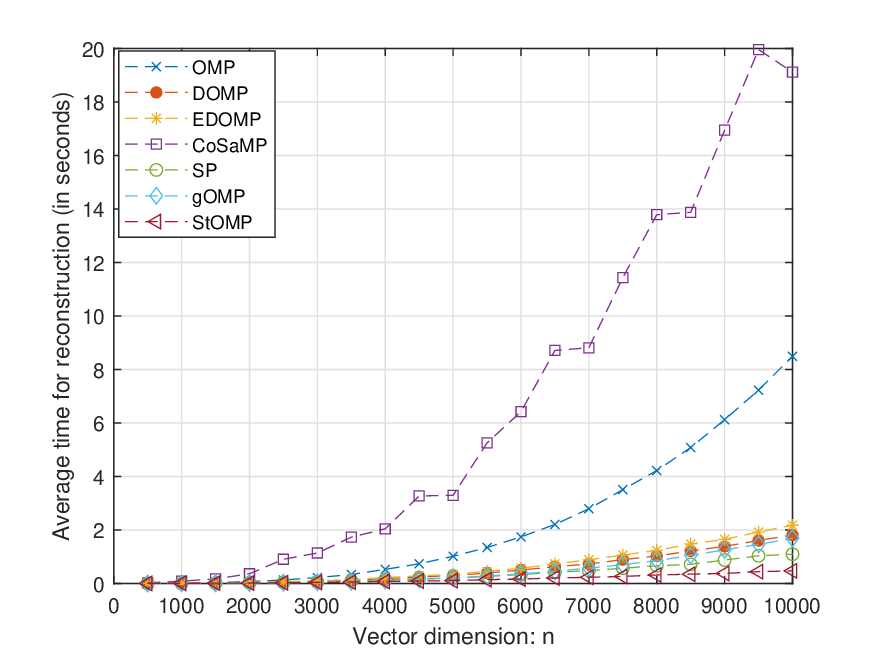} \\
\textrm{(b) Average runtime}
\end{array}
$
 \caption{Comparison of average no. of iterations and runtime required by OMP, DOMP, EDOMP, CoSaMP, SP, gOMP and StOMP} \label{figure3}
\end{figure}
The random matrices in 10 different sizes ranging from $(m,n)= (200, 1000)$ to $(2000, 10000)$ are used in this experiments. Specifically, $n=5m$ and $ m= 200 j, $ where $ j =1, \dots,10 . $ For every $m\times n$ matrix, the sparsity level of $x \in \mathbb{R}^{n}$ is set as $k=0.3m.$   The OMP algorithm always performs at least $ k $ iterations in order to possibly reconstruct the $k$-sparse vector.  Thus the total number of iterations required by OMP to reconstruct the $k$-sparse vectors ($ k=0.3m$ in this experiment) is increasing linearly with respect to the size of the matrices, as shown in Fig. \ref{figure3}(a). However, Fig. \ref{figure3}(a) indicates that  the average numbers of iterations required by DOMP and EDOMP to meet the reconstruction criterion (\ref{CRC}) are slowly increasing to the size of problems,   remarkably lower than that of CoSaMP and OMP, and very comparable to gOMP ($N=5$), but slightly more than that of SP and StOMP (with threshold parameter $t_s=3$).

 The average runtime for these algorithms to achieve the same criterion (\ref{CRC}) is summarized in Fig. \ref{figure3}(b). It can be seen that the proposed algorithms require much less computational time than OMP and CoSaMP to achieve the same reconstruction criterion. The proposed algorithms take very similar amount of time as SP, StOMP and gOMP. As the problem size goes up, the runtime of the proposed algorithms increases very slowly in a linear-like  manner, while the runtime for OMP and CoSaMP increases dramatically with respect to the size of problems.  The simulation does indicate that the total iterations as well as the runtime required by DOMP and EDOMP to reconstruct the random sparse data can be remarkably lower than that of OMP and CoSaMP.

\subsection{Comparison to some existing algorithms}
Finally, we compare the reconstruction success rates of the proposed algorithms and a few existing methods including OMP, SP CoSaMP, gOMP, StOMP and $ \ell_1$-minimization.  The parameter $ \gamma=0.9$  is still set in DOMP and EDOMP which are performed the same number of iterations as OMP and gOMP, which is set to be the sparsity level $k$ of the  target signal. In our experiments, $N=5$ is set in gOMP, and the threshold parameter $ t_s=3$ is set in StOMP.  CoSaMP and SP are allowed to perform a total of 500 iterations for every problem instance in this experiment. For accurate measurements, all algorithms use  the  stopping criterion (\ref{CRC}).  For noisy measurements, however, we adopt the following criterion: $$ \|x^{(p)}- x\|_2/\|x\|_2 \leq 10^{-3}.$$  In this comparison, the size of matrices is $500 \times 2000,$ and the sparsity level $k$ of the random data $ x \in \mathbb{R}^{2000} $ is ranged from 1 to 300 with stepsize 3, i.e., $ k =1, 4, 7, \dots, 299.$  For each given sparsity level, 200 random pairs of $(A,x)$ were realized and used to estimate the success rates of the algorithms for data reconstruction. All matrices $A$ are normalized. The accurate measurements are taken as $y: =Ax$  and inaccurate measurements are taken as $ \widetilde{y}:= Ax + 0.001h,$ where $ h$ is a normalized Gaussian random vector. When an algorithm terminates, if $x^{(p)} $ satisfies the reconstruction criterion,  a `success' is counted for the algorithm.
\begin{figure} [htp]
 $ \begin{array}{c}
\includegraphics [width=0.45\textwidth,
totalheight=0.225\textheight] {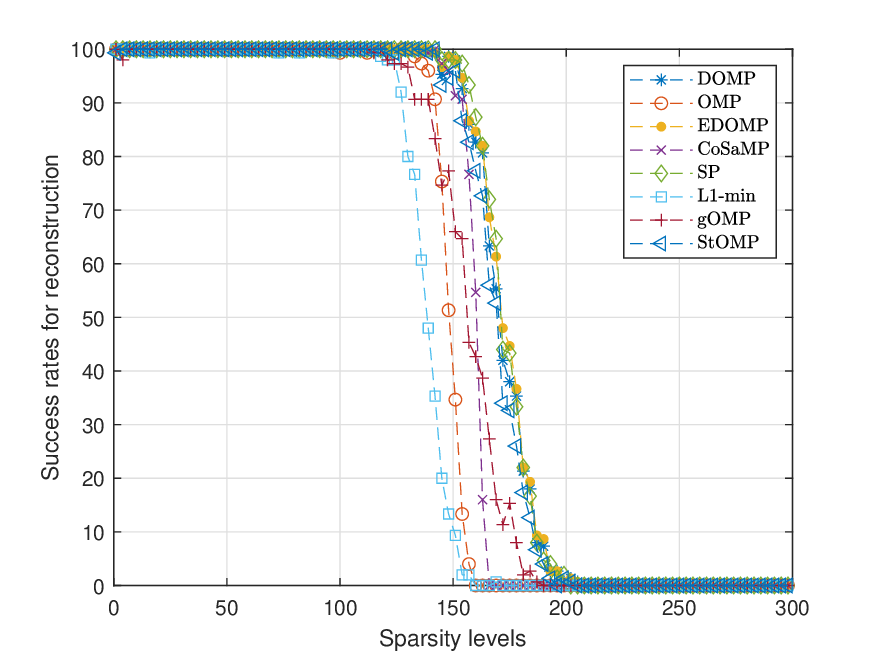} \\
  \textrm{(a) Accurate measurements }
\end{array} $
 \hfill   $\begin{array}{c}
\includegraphics [width=0.45\textwidth,
totalheight=0.225\textheight] {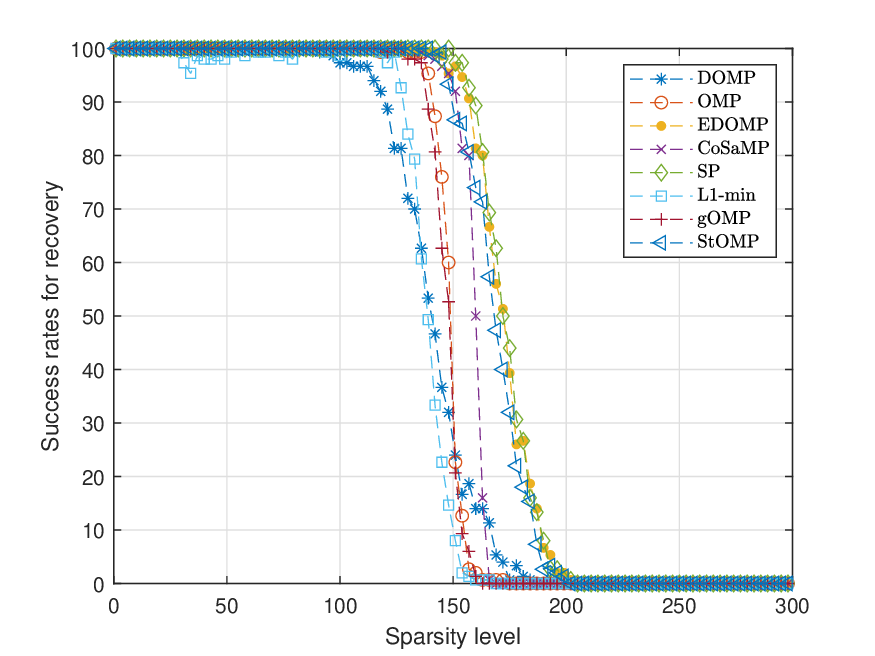} \\
\textrm{(b)  Inaccurate measurements}
\end{array}
$
 \caption{Comparison of reconstruction success rates of  algorithms in noiseless and noisy cases } \label{figure5}
\end{figure}
 The success rates of the proposed algorithms using accurate measurements are shown in Fig. \ref{figure5}(a), and the results   with inaccurate measurements are given in Fig. \ref{figure5}(b).
One can see  that in noiseless setting, DOMP and EDOMP outperform OMP, CoSaMP, gOMP and $ \ell_1$-minimization, and they perform very comparably  to SP and StOMP. The similar results for EDOMP can be observed when the measurements are slightly inaccurate, while DOMP and gOMP are clearly subject to the effect of noises.
In summary,  EDOMP is very competitive to the existing mainstream algorithms for sparse data reconstruction. It is fast, stable, robust and very easy to implement like CoSaMP and SP.

\section{Conclusions and future work}\label{sect6}
The reconstruction error bound for the proposed DOMP algorithm  is established in this paper under the RIP assumption. Finite convergence of the algorithm is also discussed in large-scale compressed sensing scenarios. Simulations indicate that the proposed methods (especially, EDOMP) are more efficient than OMP in sense that they may take much less average computational times and iterations to reconstruct the sparse data, and they can compete against several state-of-art methods in this field.  Distinguished from the existing OMP-type methods, the main feature of the proposed algorithms is that they can dynamically  and efficiently exploit the gradient information of the error metric at every iteration to recognize the support of the target data.

 \end{document}